\documentclass[journal]{IEEEtran}
\usepackage{cite}
\usepackage{amsmath}
\usepackage{amsfonts}
\usepackage{color}
\usepackage{amssymb}
\usepackage{algorithm}
\usepackage{algorithmic}
\usepackage{array}
\usepackage{url}
\usepackage{graphicx}
\usepackage{subfigure}
\usepackage{longtable}
\usepackage{multirow}
\usepackage{marvosym}
\usepackage{textcomp}
\usepackage{threeparttable, tablefootnote}
\usepackage{enumerate}
\usepackage[justification=centering]{caption}

\usepackage{xcolor}
\usepackage{soul}
\soulregister\cite7
\soulregister\ref7

\allowdisplaybreaks[4]
\renewcommand{\arraystretch}{1.2}
\newtheorem{lemma}{Lemma}

\newtheorem{theorem}{Theorem}

\begin{document}
%
\title{PORA: Predictive Offloading and Resource Allocation in Dynamic Fog Computing Systems}
%
%
%

\author{Xin~Gao,~\IEEEmembership{Student Member,~IEEE,}
		Xi~Huang,~\IEEEmembership{Student Member,~IEEE,}
		Simeng~Bian,~\IEEEmembership{Student Member,~IEEE,}
        Ziyu~Shao$^{*}$,~\IEEEmembership{Member,~IEEE,}
        Yang~Yang,~\IEEEmembership{Fellow,~IEEE}
\thanks{
X. Gao, X. Huang, S. Bian, Z. Shao and Y. Yang are with the School of Information Science and Technology, ShanghaiTech University, Shanghai 201210, China. (E-mail: \{gaoxin, huangxi, biansm, shaozy, yangyang\}@shanghaitech.edu.cn)
(*Corresponding author: Ziyu Shao)
}}
\maketitle

\begin{abstract}
In multi-tiered fog computing systems, to accelerate the processing of computation-intensive tasks for real-time IoT applications, 
resource-limited IoT devices can offload part of their workloads to nearby fog nodes, 
whereafter such workloads may be offloaded to upper-tier fog nodes with greater computation capacities. 
Such hierarchical offloading, though promising to shorten processing latencies, may also induce excessive power consumptions and latencies for wireless transmissions. 
With the temporal variation of various system dynamics, such a trade-off makes it rather challenging to conduct effective and online offloading decision making.
Meanwhile, the fundamental benefits of predictive offloading to fog computing systems still remain unexplored.
In this paper, we focus on the problem of dynamic offloading and resource allocation with traffic prediction in multi-tiered fog computing systems.  
By formulating the problem as a stochastic network optimization problem, we aim to minimize the time-average power consumptions with stability guarantee for all queues in the system. 
We exploit unique problem structures and propose PORA, an efficient and distributed predictive offloading and resource allocation scheme for multi-tiered fog computing systems.  
Our theoretical analysis and simulation results show that PORA incurs near-optimal power consumptions with queue stability guarantee. 
Furthermore, PORA requires only mild-value of predictive information to achieve a notable latency reduction, even with prediction errors.  
\end{abstract}


%
\IEEEpeerreviewmaketitle

\section{Introduction}

In the face of the proliferation of real-time IoT applications, fog computing has come as a promising complement to cloud computing by extending cloud to the edge of the network to meet the stringent latency requirements and intensive computation demands of such applications \cite{xiao2017qoe}.

A typical fog computing system consists of a set of geographically distributed fog nodes
which are deployed at the network periphery with elastic resource provisioning such as storage, computation, and network bandwidth\cite{yi2015fog}. 
Depending on their distance to IoT devices, fog nodes are often organized in a hierarchical fashion, with each layer as a \textit{fog tier}.
In such a way, resource-limited IoT devices, when heavily loaded, can delegate workloads via wireless links to nearby fog nodes, \textit{a.k.a.}, \textit{workload offloading}, 
to reduce the power consumption and accelerate workload processing; 
meanwhile, each fog node can offload workloads to nodes in its upper fog tier. 
However, along with all the benefits come the extended latency and extra power consumption.   
Given such a power-latency tradeoff, two interesting questions arise.
One is \textit{where} and \textit{how much workloads} to offload between successive fog tiers. 
The other is how to \textit{allocate resources} for workload processing and offloading.
The timely decision making regarding these two questions is critical but challenging, due to temporal variations of system dynamics in wireless environment, uncertainty in the resulting offloading latency, and the unknown traffic statistics.

We summarize the main challenges of dynamic offloading and resource allocation in fog computing as follows: 
\begin{enumerate}
	\item[$\diamond$] \textbf{Characterization of system dynamics and the power-latency tradeoff}: In practice, a fog system often consists of multiple tiers, with complex interplays between fog tiers and the cloud, not to mention the constantly varying dynamics and intertwined power-latency tradeoffs therein.
		A model that accurately characterizes the system and tradeoffs is the key to the fundamental understanding of the design space.
	\item[$\diamond$] \textbf{Efficient online decision making:}
		The decision making must be computationally efficient, so as to minimize the overheads. The difficulties often come from the uncertainties of traffic statistics, online nature of workload arrivals, and intrinsic complexity of the problem.
	\item[$\diamond$] \textbf{Understanding the benefits of predictive offloading:} One natural extension to online decision making is to employ predictive offloading to further reduce latencies and improve quality of service. For example, Netflix preloads videos onto users' devices based on user behavior prediction\cite{NetflixPred}. Despite the wide applications of such approaches, the fundamental limits of predictive offloading in fog computing still remain unknown.
\end{enumerate}

\begin{table*}[!h]
\centering
\caption{Comparisons of related works}
\label{table: related works}
\begin{threeparttable}	
\begin{tabular}{|c|c|c|c|c|c|c|c|}
\hline
 & D2D-enabled IoT & IoT-Fog\tnote{1} & Fog-Fog\tnote{2} & Fog-Cloud\tnote{3} & Dynamic & Prior Arrival Distribution & Prediction \\
\hline
\cite{xiao2017qoe} &  & \checkmark & \checkmark & \checkmark &  & -- &  \\
\hline
\cite{wang2019cooperative} &  & \checkmark &  & \checkmark &  & -- &  \\
\hline
\cite{liu2018socially} &  & \checkmark &  & \checkmark & \checkmark & Poisson & \\
\hline 
\cite{misra2019detour} &  & \checkmark &  &  &  & -- &  \\
\hline
\cite{lei2019joint} &  & \checkmark &  &  & \checkmark & Poisson &  \\
\hline
\cite{mao2016power} &  & \checkmark &  &  & \checkmark & Not Required &  \\
\hline
\cite{chen2019energy} &  & \checkmark &  &  & \checkmark & Not Required &  \\
\hline
\cite{gao2019dynamic} & \checkmark & \checkmark &  &  & \checkmark & Not Required &  \\
\hline
\cite{zhang2019near} &  & \checkmark &  &  & \checkmark & Not Required &  \\
\hline
Ours &  &  & \checkmark & \checkmark & \checkmark &  Not Required & \checkmark \\
\hline
\end{tabular}
\begin{tablenotes}
	\footnotesize
	 \item[1,2,3] ``IoT-Fog'' means offloading from IoT devices to fog, ``Fog-Fog'' means offloading between fog tiers, while ``Fog-Cloud'' means offloading from fog to cloud. 
\end{tablenotes}
\end{threeparttable}
	\vspace{-0.5cm}  
\end{table*}

In this paper, we focus on the workload offloading problem for multi-tiered fog systems. We address the above challenges by developing a fine-grained queueing model that accurately depicts such systems and proposing an efficient online scheme that proceeds the offloading on a time-slot basis. 
To the best of our knowledge, we are the first to conduct systematic study on predictive offloading in fog systems. 
Our key results and main contributions are summarized as follows:
\begin{enumerate}
	\item[$\diamond$] \textbf{Problem Formulation:} We formulate the problem of dynamic offloading and resource allocation as a stochastic optimization problem, aiming at minimizing the long-term time-average expectation of total power consumptions of fog tiers with queue stability guarantee.
	\item[$\diamond$] \textbf{Algorithm Design:} 
		Through a non-trivial transformation, 
		we decouple the problem into a series of subproblems over time slots. By exploiting their unique structures, we propose PORA, an efficient scheme that exploits predictive scheduling to make decisions in an online manner.
	\item[$\diamond$] \textbf{Theoretical Analysis and Experimental Verification:} We conduct theoretical analysis and trace-driven simulations to evaluate the effectiveness of PORA. 
		The results show that PORA achieves a tunable power-latency tradeoff while effectively reducing the average latency with only mild-value of predictive information, even in the presence of prediction errors.
	\item[$\diamond$] \textbf{New Degree of Freedom in the Design of Fog Computing Systems:} We systematically investigate the fundamental benefits of predictive offloading in fog computing systems, with both theoretical analysis and numerical evaluations.
\end{enumerate}


We organize the rest of the paper as follows. 
Section \ref{sec: related work} discusses the related work. 
Next, in Section \ref{sec: motivating example}, we provide an example that motivates our design for dynamic offloading and resource consumption in fog computing systems. 
Section \ref{sec: model} presents the system model and problem formulation, followed by the algorithm design of PORA and performance analysis in Section \ref{sec: algorithm}.
Section \ref{sec: simulation} analyzes the results from trace-driven simulations, while Section \ref{sec: conclusion} concludes the paper.

\section{Related Work}\label{sec: related work}

In recent years, a series of works have been proposed to optimize the performance fog computing systems from various aspects \cite{xiao2017qoe, taneja2017resource, chen2019dynamic, wang2019cooperative, liu2018socially, misra2019detour, lei2019joint, mao2016power, chen2019energy, gao2019dynamic, zhang2019near}.
Among such works, the most related are those focusing on the design of effective offloading schemes. 
For example, 
by adopting alternating direction method of multipliers (ADMM) methods,
Xiao \textit{et al.}\cite{xiao2017qoe} and Wang \textit{et al.}\cite{wang2019cooperative} proposed two offloading schemes for cloud-aided fog computing systems to minimize average task duration and average service response time under different energy constraints, respectively. 
Later, Liu \textit{et al.} \cite{liu2018socially} took the social relationships among IoT users into consideration and developed a socially aware offloading scheme by advocating game theoretic approaches. 
Misra \textit{et al.} \cite{misra2019detour} studied the problem in software-defined fog computing systems and proposed a greedy heuristic scheme to conduct multi-hop task offloading with offloading path selection. 
Lei \textit{et al.} \cite{lei2019joint} considered the joint minimization of delay and power consumption over all IoT devices; they formulated the problem under the settings of continuous-time Markov decision process and solved it via approximate dynamic programming techniques.
The above works, despite their effectiveness, generally assume the availability of the statistical information on task arrivals in the systems which is usually unattainable in practice with highly time-varying system dynamics \cite{zhang2017resource}.

In the face of such uncertainties, a number of works have applied stochastic optimization methods such as Lyapunov optimization techniques to online and dynamic offloading scheme design 
\cite{mao2016power, chen2019energy, gao2019dynamic, zhang2019near}. 
For instance, Mao \textit{et al.}\cite{mao2016power} investigated the tradeoff between the power consumption and execution delay, then developed a dynamic offloading scheme for energy-harvesting-enabled IoT devices.
Chen \textit{et al.} \cite{chen2019energy} designed an adaptive and efficient offloading scheme to minimize the transmission energy consumption with queueing latency guarantee.
Gao \textit{et al.} \cite{gao2019dynamic} investigated efficient offloading and social-awareness-aided network resource allocation for device-to-device-enabled (D2D-enabled) IoT users. 
Zhang \textit{et al.} \cite{zhang2019near} designed an online rewards-optimal scheme for the computation offloading of energy harvesting-enabled IoT devices based on Lyapunov optimization and Vickrey-Clarke-Groves auction.
Different from such works that focus on fog computing systems with flat or two-tiered architectures, 
our solution is applicable to general multi-tiered fog computing systems with time-varying wireless channel states and unknown traffic statistics. 
Moreover, to the best of our knowledge, 
our solution is also the first to proactively leverage the predicted traffic information to optimize the system performance with theoretical guarantee. 
We are also the first to investigate the fundamental benefits of predictive offloading in fog computing systems.
We compare our work with the above mentioned works 
in TABLE \ref{table: related works}.

\section{Motivating Example}\label{sec: motivating example}

In this section, we provide a motivating example to show the potential power-latency tradeoff in multi-tiered fog computing systems. 
The objective is to achieve low power consumptions and short average workload latency (in the unit of packets). 

\begin{figure}[!t]
	\centering
	\includegraphics[scale=0.44]{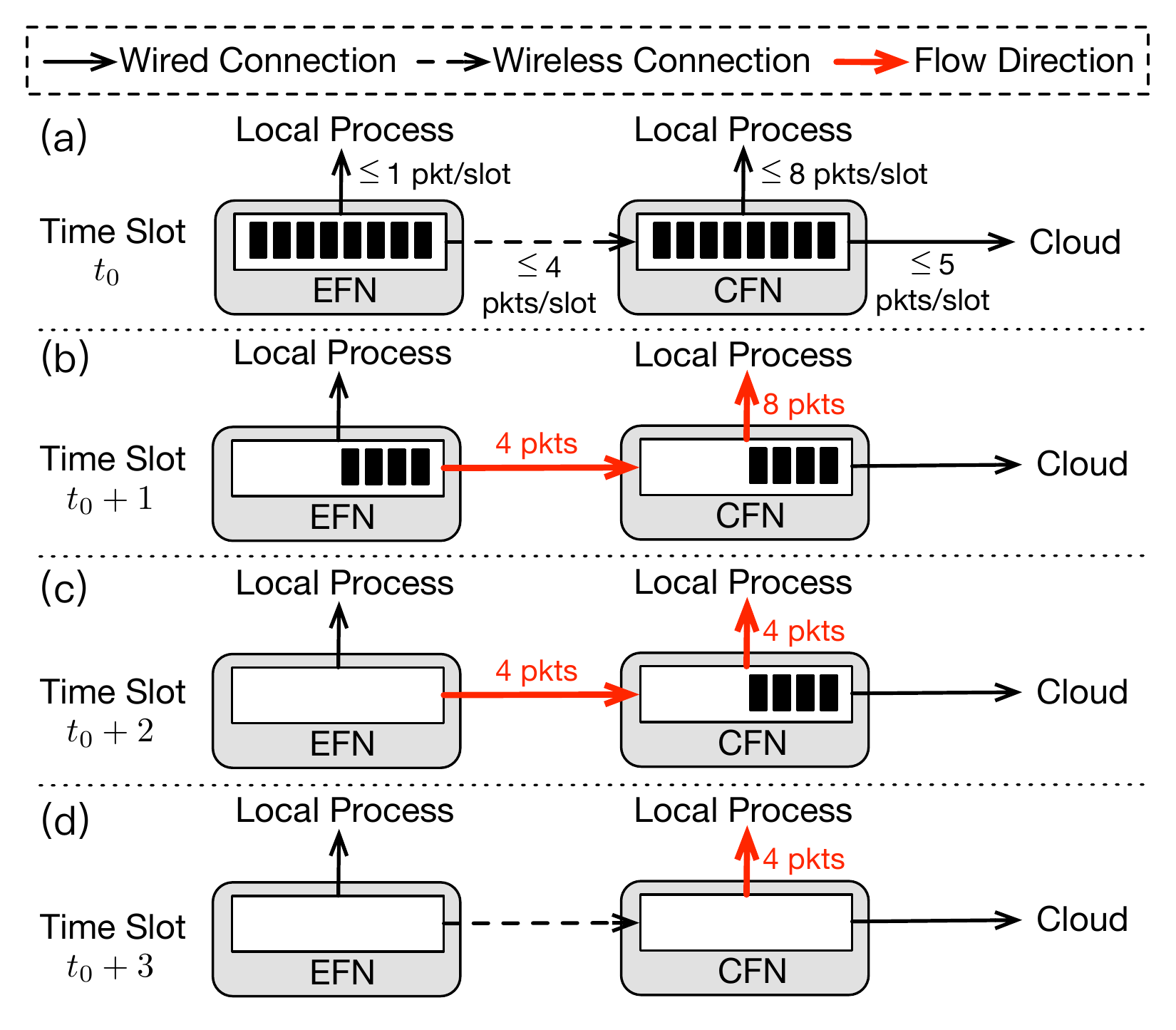}
	\caption{Motivating example of dynamic offloading and resource consumption in multi-tiered fog computing systems.}
	\label{figure: motivating example}
\end{figure}

Figure \ref{figure: motivating example} shows an instance of time-slotted fog computing system with two fog tiers, \textit{i.e.}, edge fog tier and central fog tier. 
Within each fog tier resides one fog node, \textit{i.e.}, an edge fog node (EFN) in edge fog tier and a central fog node (CFN) in central fog tier. 
The EFN connects to the CFN via a wireless link, while the CFN connects to the cloud data center over wired links. 
Each fog node maintains one queue to store packets. 
Figure \ref{figure: motivating example}(a) shows that during time slot $t_{0}$, both the EFN and the CFN store $8$ packets in their queues. 

We assume that each fog node sticks to one policy all the time to handle packets, \textit{i.e.}, either \textit{processing packets locally} or \textit{offloading them to its next tier}.
The local processing capacities of EFN and CFN are $1$ and $8$ packets per time slot, respectively.
The transmission capacities from EFN to CFN and from CFN to cloud are $4$ and $5$ packets per time slot, respectively. 
The power consumption is assumed linearly proportional to the number of processed/transmitted packets. 
In particular, processing one packet locally consumes $1$ mW power, while transmitting one packet over wireless link consumes $0.5$ mW. 
We ignore the processing latency in the cloud due to its powerful processing capacity.  

TABLE \ref{table: motivating example} lists the total power consumptions and average packet latencies under all four possible settings. 
Figures \ref{figure: motivating example}(b)-\ref{figure: motivating example}(d) show the case when EFN sticks to offloading and CFN sticks to local processing. 
In time slot $(t_{0}+1)$, EFN offloads four packets to CFN at its full transmission capacity, while CFN processes all the eight packets locally. 
In time slot $(t_{0}+2)$, EFN offloads the rest four packets to CFN; meanwhile, CFN locally processes the four packets that arrive in previous time slot. 
In time slot $(t_{0}+3)$, CFN finishes processing the rest four packets. 
In this case, the system consumes $16$ mW power in local processing and $4$ mW power in transmission, with an average packet latency of $1.75$ time slots. 

\begin{table}[!h]
\centering
\caption{Performance under different offloading policies}
\label{table: motivating example}
\begin{tabular}{|c|c|c|c|}
\hline
Policy of & Policy of & Total Power & Average Packet \\ 
EFN & CFN & Consumptions (mW) & Latency (time slot) \\ \hline
Local & Local & 16 & 2.75 \\ \hline
Local & Offload & 8 & 2.9375 \\ \hline
Offload & Local & 20 & 1.75 \\ \hline
Offload & Offload & 4 & 2.125 \\ \hline
\end{tabular}
\end{table}

From TABLE \ref{table: motivating example}, we conclude that: 
First, when EFN sticks to offloading and CFN sticks to local processing, the system achieves the lowest average packet latency of $1.75$ slots but the maximum power consumption of $20$mW. 
Second, with the same offloading policy on EFN, there is a tradeoff between the total power consumptions and the average packet latency when CFN sticks to different policies. 
The reason is that offloading to the cloud can not only reduce power consumptions but  also prolong latency as well. 
Third, when CFN sticks to local processing, there is a power-latency tradeoff with different policies at EFN, in that offloading to CFN can induce lower processing latency but at the cost of even higher power consumption for wireless transmissions.

\section{Model and Problem Formulation}
\label{sec: model}

\begin{figure}[!t]
	\centering
	\includegraphics[scale=0.35]{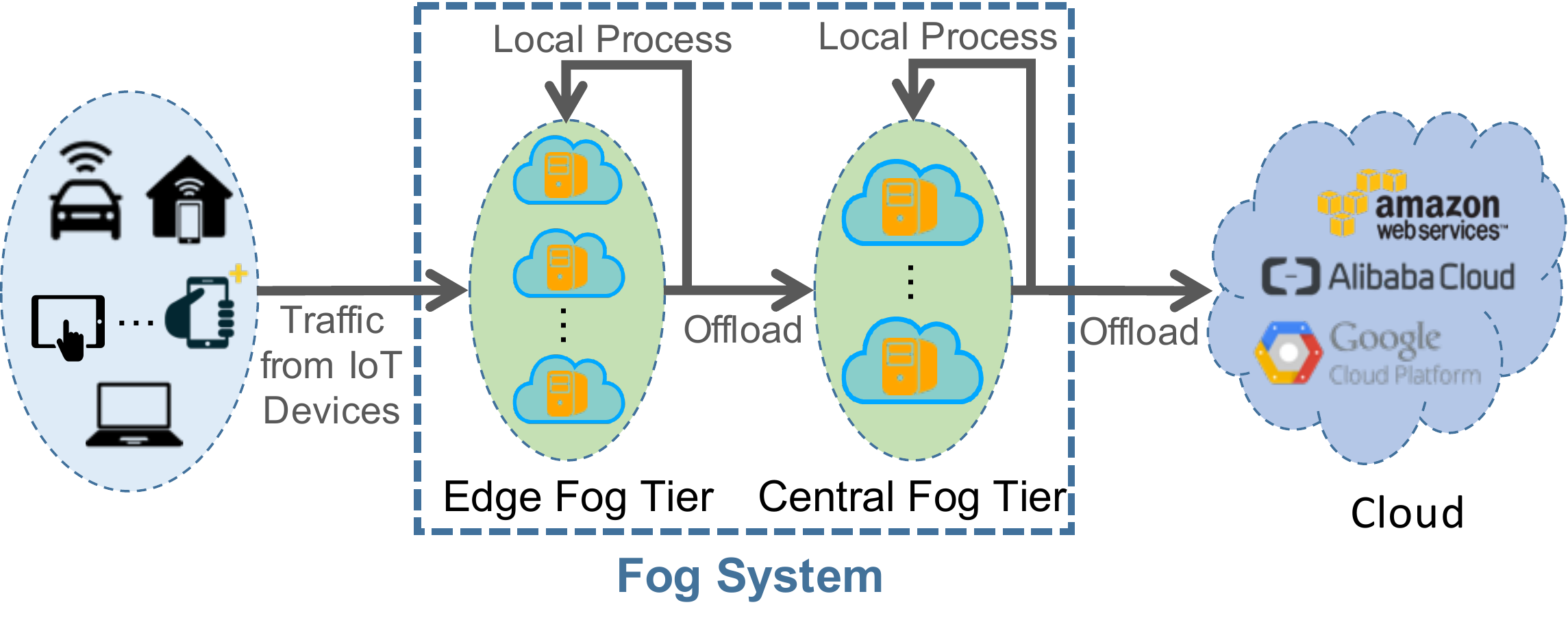}
	\caption{An example of fog computing systems with two fog tiers.}
	\label{figure: hierarchical fog system}
\end{figure}

We consider a multi-tiered fog computing system, as shown in Figure \ref{figure: hierarchical fog system}.
The system evolves over time slots indexed by $t \in \{0, 1, 2,...\}$.
Each time slot has a length of $\tau_{0}$.
Inside the edge fog tier (EFT) are a set of edge fog nodes (EFNs) that offer low-latency access to IoT devices.
On the other hand, the central fog tier (CFT) comprises of central fog nodes (CFNs) with greater processing capacities than EFNs.
We assume that the workload on each EFN can be offloaded to and processed by any of its accessible CFNs, and that each CFN can offload its workload to the cloud.
In our model, we do not consider the power consumptions and latencies within the cloud. We mainly focus on the power consumptions and latencies within fog tiers, as shown in TABLE \ref{table: model}. First, the power consumptions we consider include two parts: processing power and transmit power. The processing power consumption is induced by the workload processing on both EFT and CFT. The transmit power is induced by the transmissions from EFT to CFT. We do not consider the transmit power consumption from CFT to cloud because we assume that the CFT communicates with the cloud through wireline connections. Second, the latencies we consider include three parts: queueing latency, processing latency and transmit latency. We focus on the queueing latency on both EFT and CFT. We assume that the workload processing in each time slot can be completed by the end of the same time slot, and then we can ignore the processing latency. Since the EFT communicates with the  CFT through high-speed  wireless connections and the  CFT communicates with the cloud through high-speed wireline connections, we assume that transmission latencies from both EFT to CFT  and CFT to Cloud are negligible.

\begin{table}[!h]
\centering
\caption{Performance Metrics in Our Model}
\label{table: model}
\begin{tabular}{|p{1.2cm}<{\centering}|p{1.1cm}<{\centering}|p{0.9cm}<{\centering}|p{1.0cm}<{\centering}|p{1.1cm}<{\centering}|p{0.9cm}<{\centering}|}
\hline
\multirow{2}{*}{} & \multicolumn{2}{c|}{Power Consumption} & \multicolumn{3}{c|}{Latency} \\ \cline{2-6} 
& Processing & Transmit & Queueing & Processing & Transmit \\ \hline
EFT & \checkmark &  & \checkmark & &  \\ \hline
EFT2CFT &  & \checkmark &  &  &  \\ \hline
CFT & \checkmark &  & \checkmark & &  \\ \hline
CFT2Cloud &  &  &  &  &  \\ \hline
\end{tabular}
\end{table}

In the following, we first introduce the basic settings in Section \ref{subsec: basic setting}, then elaborate the queueing models in Section \ref{subsec: local queue}. 
Next, we define the optimization objective in Section \ref{subsec: power} while proposing the problem formulation in Section \ref{subsec: formulation}.
We summarize the key notations in TABLE \ref{table: key notations}.

\begin{table}[!t]
\renewcommand{\arraystretch}{1.3}
\caption{Key notations}
\label{table: key notations}
\centering
\begin{tabular}{p{0.89cm} l}
    \hline\hline
    Notation & Description  \\ \hline
    $\tau_{0}$ & Length of each time slot  \\ \hline
    $\mathcal{N}$ & $\mathcal{N}$ is the set of EFNs with $|\mathcal{N}|\triangleq N$ \\ \hline
    $\mathcal{M}$ & $\mathcal{M}$ is the set of CFNs with $|\mathcal{M}|\triangleq M$  \\ \hline
    $\mathcal{N}_{j}$ & Set of accessible EFNs from CFN $j$  \\ \hline    
    $\mathcal{M}_{i}$ & Set of accessible CFNs from EFN $i$  \\ \hline
	$A_{i}(t)$ & Amount of workload arriving to EFN $i$ in time slot $t$  \\ \hline
	$\lambda_{i}$ & Average workload arriving rate on EFN $i$, $\lambda_{i}\triangleq \mathbb{E}\{A_{i}(t)\}$  \\ \hline
	$W_{i}$ & Prediction window size of EFN $i$  \\ \hline
	$A_{i,-1}(t)$ & Arrival queue backlog of EFN $i$ in time slot $t$  \\ \hline
	\multirow{2}*{$A_{i,w}(t)$} & Prediction queue backlog of EFN $i$ in time slot $t$, such that  \\ 
	& $0\leq w\leq W_{i}-1$  \\ \hline
	$Q_{i}^{(e,a)}(t)$ & Integrate queue backlog of EFN $i$ in time slot $t$  \\ \hline
	$Q_{i}^{(e,l)}(t)$ & Local processing queue backlog of EFN $i$ in time slot $t$  \\ \hline
	$Q_{i}^{(e,o)}(t)$ & Offloading queue backlog of EFN $i$ in time slot $t$  \\ \hline
	$b_{i}^{(e,l)}(t)$ & Amount of workload to be sent to $Q_{i}^{(e,l)}(t)$ in time slot $t$ \\ \hline
	$b_{i}^{(e,o)}(t)$ & Amount of workload to be sent to $Q_{i}^{(e,o)}(t)$ in time slot $t$ \\ \hline
	$f_{i}^{(e)}(t)$ & CPU frequency of EFN $i$ in time slot $t$ \\ \hline
	$H_{i,j}(t)$ & Wireless channel gain between EFN $i$ and CFN $j$ \\ \hline
	$p_{i,j}(t)$ & Transmit power from EFN $i$ to CFN $j$ in time slot $t$ \\ \hline
	$R_{i,j}(t)$ & Transmit rate from EFN $i$ to CFN $j$ in time slot $t$ \\ \hline
	$Q_{j}^{(c,a)}(t)$ & Arrival queue backlog of CFN $j$ in time slot $t$  \\ \hline
	$Q_{j}^{(c,l)}(t)$ & Local processing queue backlog of CFN $j$ in time slot $t$  \\ \hline
	$Q_{j}^{(c,o)}(t)$ & Offloading queue backlog of CFN $j$ in time slot $t$  \\ \hline
	$b_{j}^{(c,l)}(t)$ & Amount of workload to be sent to $Q_{j}^{(c,l)}(t)$ in time slot $t$ \\ \hline
	$b_{j}^{(c,o)}(t)$ & Amount of workload to be sent to $Q_{j}^{(c,o)}(t)$ in time slot $t$ \\ \hline
	$f_{j}^{(c)}(t)$ & CPU frequency of CFN $j$ in time slot $t$ \\ \hline
	$P(t)$ & Total power consumptions in time slot $t$ \\ \hline
	\hline
\end{tabular}
\end{table}

\subsection{Basic Settings}\label{subsec: basic setting}

The fog computing system consists of $N$ EFNs in EFT and $M$ CFNs in CFT. 
Let $\mathcal{N}$ and $\mathcal{M}$ be the sets of EFNs and CFNs.
Each EFN $i$ has access to a subset of CFNs in their proximities. We denote the subset by $\mathcal{M}_{i}\subset\mathcal{M}$. 
For each CFN $j$, $\mathcal{N}_{j}\subset\mathcal{N}$ denotes the set of its accessible EFNs. 
Accordingly, for any $i\in\mathcal{N}_{j}$ we have $j\in\mathcal{M}_{i}$. 

\subsection{Queueing Model for Edge Fog Node}\label{subsec: queue for EFN}

During time slot $t$, there is an amount $A_{i}(t)$ ($\le A_{\text{max}}$ for some constant $A_{\text{max}}$) of workload generated from IoT devices arrive to be processed on EFN $i$ such that $\mathbb{E}\{ A_{i}(t)\} =\lambda_{i}$.
We assume that such arrivals are independent over time slots and different EFNs. 
Each EFN $i$ is equipped with a learning module\footnote{We do not specify any particular learning method in this paper, since our work aims to explore the \textit{fundamental} benefits of predictive offloading. 
In practice, one can leverage machine learning techniques such as time-series prediction methods \cite{ahmed2010empirical} for workload arrival prediction.}
that can predict the future workload within a \textit{prediction window} of size $W_{i}$, \textit{i.e.} workload will arrive in the next $W_{i}$ time slots. The predicted arrivals are pre-generated and recorded, then arrive to EFN $i$ for pre-serving. Once the predicted arrivals actually arrive after pre-serving, they will be considered finished.

\begin{figure}[!t]
	\centering
	\includegraphics[scale=0.3]{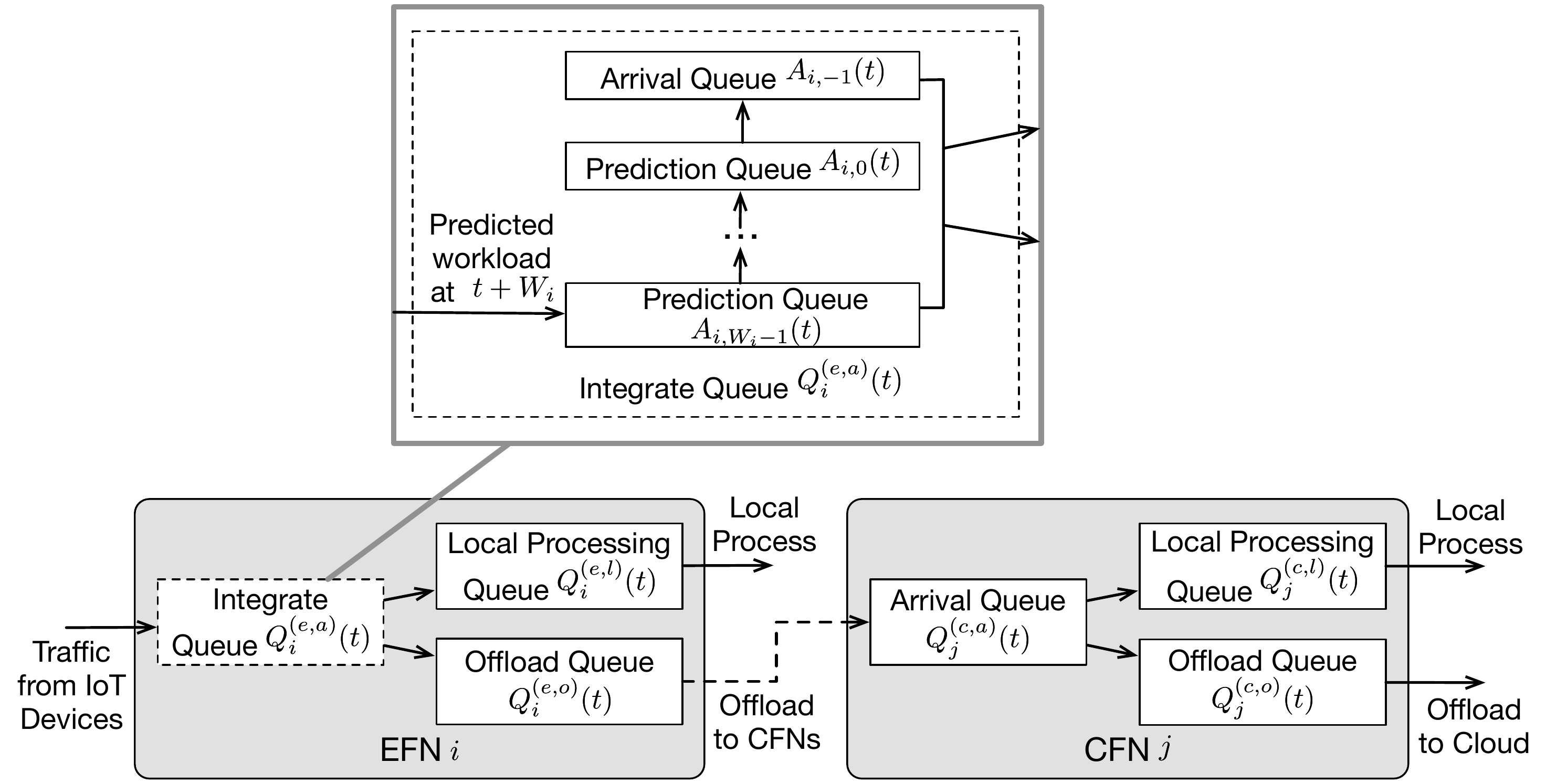}
	\caption{Queueing model of the system.}
	\label{figure: queue}
\end{figure}

On each EFN, as Figure \ref{figure: queue} shows, there are four types of queues:
prediction queues with the backlogs as $A_{i,0}(t)$, ..., $A_{i,W_{i}-1}(t)$, 
arrival queue $A_{i,-1}(t)$, 
local processing queue $Q_{i}^{(e,l)}(t)$, and offloading queue $Q_{i}^{(e,o)}(t)$.
In time slot $t$, prediction queue $A_{i,w}(t)$ ($0\leq w\leq W_{i}-1$) stores untreated workload that will arrive in time slot $(t+w)$. 
Workload that actually arrives at EFN $i$ is stored in the arrival queue $A_{i,-1}(t)$, 
awaiting being forwarded to the local processing queue $Q_{i}^{(e,l)}(t)$ or the offloading queue $Q_{i}^{(e,o)}(t)$. 
Workload in $Q_{i}^{(e,l)}(t)$ will be processed locally by EFN $i$, 
while workload in $Q_{i}^{(e,o)}(t)$ will be offloaded to CFNs in set $\mathcal{M}_{i}$.

\subsubsection{Prediction Queues and Arrival Queues in EFNs}

Within each time slot $t$, in addition to the current arrivals in the arrival queue, EFN $i$ can also forward future arrivals in the prediction queues.
We define $\mu_{i,w}(t)$ as the amount of output workload from $A_{i,w}(t)$, for $w\in\{-1,0,...,W_{i}-1\}$.
Such workload should be distributed to the local processing queue and offloading queue.
We denote the amounts of workloads to be distributed to the local processing queue and offloading queue as $b_{i}^{(e,l)}(t)$ and $b_{i}^{(e,o)}(t)$, respectively, such that
\begin{equation}
	0\leq b^{(e,\beta)}_{i}\left(t\right)\leq b^{(e,\beta)}_{i,\text{max}},\ \forall \beta\in\{l,o\} \label{constraint: EFN offloading decision 1}
\end{equation}
where each $b^{(e,\beta)}_{i,\text{max}}$ is a positive constant.
As a result, we have
\begin{equation}\label{constraint: pre-serve rates}
	\sum_{w=-1}^{W_{i}-1}\mu_{i,w}\left(t\right)= b^{(e,l)}_{i}\left(t\right)+b^{(e,o)}_{i}\left(t\right).
\end{equation}

Next, we consider the queueing dynamics for different types of queues in EFN, respectively.

Regarding $A_{i,w}(t)$, it is updated whenever pre-service is finished and the lookahead window moves one slot ahead at the end of each time slot. 
Therefore, we have
\begin{enumerate}[(i)]
	\item If $w=W_{i}-1$, then
	\begin{equation}\label{update: fog node prediction queue update 1}
		A_{i,W_{i}-1}\left(t+1\right)=A_{i}\left(t+W_{i}\right).
	\end{equation}
	\item If $0\leq w\leq W_{i}-2$, then
	\begin{equation}\label{update: fog node prediction queue update 2}
		A_{i,w}(t+1)=[A_{i,w+1}(t)-\mu_{i,w+1}(t)]^{+},
	\end{equation}
\end{enumerate}
where $[x]^{+}\triangleq \max\{x,0\}$ for $x \in \mathbb{R}$. 
In time slot $(t+1)$, the amount of workload that will arrive after $(W_{i}-1)$ time slots is $A_{i}(t+W_{i})$ and it remains unknown until time slot $(t+1)$. 

Regarding the arrival queue $A_{i,-1}(t)$, it records the actual backlog of EFN $i$ with the update equation as follows:
\begin{equation}\label{update: fog node true queue update}
	A_{i,-1}(t+1)\!=\![A_{i,-1}(t)-\mu_{i,-1}(t)]^{+}\!+\![A_{i,0}(t)-\mu_{i,0}(t)]^{+}.
\end{equation}
Note that $\mu_{i,-1}(t)$ denotes the amount of distributed workload that have already being in $A_{i,-1}(t)$. 

Next, we introduce an integrate queue with a backlog size as the sum of all prediction queues and the arrival queue on EFN $i$, denoted by $Q_{i}^{\left(e,a\right)}\left(t\right)\triangleq \sum_{w=-1}^{W_{i}-1}A_{i,w}\left(t\right)$ . 
Under \textit{fully-efficient} \cite{huang2016predictive} service policy,  $Q_{i}^{\left(e,a\right)}\left(t\right)$ is updated as
\begin{multline}\label{update: prediction sum queue update}
	Q_{i}^{\left(e,a\right)}\left(t+1\right)=[Q_{i}^{\left(e,a\right)}\left(t\right)-(b^{(e,l)}_{i}\left(t\right)+b^{(e,o)}_{i}\left(t\right))]^{+}\\
	+A_{i}\left(t+W_{i}\right).
\end{multline}
The input of integrate queue $Q_{i}^{(e,a)}(t)$ consists of the predicted workload that will arrive at EFN $i$ in time slot $(t+W_{i})$, while its output consists of workloads being forwarded to the local processing queue and the offloading queue. 
Note that $b^{(e,l)}_{i}(t)+b^{(e,o)}_{i}(t)$ is the output capacity of integrate queue $Q_{i}^{(e,a)}(t)$ in time slot $t$. If the capacity is larger than the queue backlog size, the true output amount will be smaller than $b^{(e,l)}_{i}(t)+b^{(e,o)}_{i}(t)$.

\subsubsection{Offloading Queues in EFNs}

In time slot $t$, workload in queue $Q^{(e,o)}_{i}(t)$ will be offloaded to CFNs in set $\mathcal{M}_{i}$. 
The transmission capacities are determined by the transmit power decisions $(p_{i,j}(t))_{j\in\mathcal{M}_{i}}$, where $p_{i,j}(t)$ is the transmit power from EFN $i$ to CFN $j$. 
The transmit power is nonnegative and the total transmit power of each EFN is upper bounded, \textit{i.e.}, 
\begin{align}
	&p_{i,j}\left(t\right)\geq0,\ \forall i\in\mathcal{N},j\in\mathcal{M}_{i}\text{ and }t, \label{constraint: power allocation 1} \\
	&\sum_{j\in\mathcal{M}_{i} }p_{i,j}\left(t\right)\leq p_{i,\text{max}},\ \forall i\in\mathcal{N}\text{ and }t.\label{constraint: power allocation 2}
\end{align}
According to Shannon's capacity formula \cite{gallager2008principles}, the transmission capacity from EFN $i$ to CFN $j$ is
\begin{equation}\label{equation: offload rate to central fog}
	R_{i,j}(t)\!\triangleq\!\hat{R}_{i,j}(p_{i,j}(t))\!=\!\tau_{0}B\log_{2}\left(1\!+\!\frac{p_{i,j}(t)H_{i,j}(t)}{N_{0}B}\right),
\end{equation}
where $\tau_{0}$ is the length of each time slot, $B$ is the channel bandwidth, $H_{i,j}(t)$ is the wireless channel gain between EFN $i$ and CFN $j$, and $N_{0}$ is the system power spectral density of the additive white Gaussian noise.
Note that $H_{i,j}(t)$ is an uncontrollable environment state with positive upper bound $H_{\text{max}}$.
We do not consider the interferences among fog nodes and tiers. 
By adjusting the transmit power $p_{i,j}(t)$, we can offload different amounts of workload from EFN $i$ to CFN $j$ in time slot $t$. 
Accordingly, the update equation of offloading queue $Q^{(e,o)}_{i}(t)$ is
\begin{equation}\label{update: edge offloading queue update}
	Q^{(e,o)}_{i}\left(t+1\right)\leq [Q^{(e,o)}_{i}\left(t\right)\!-\!\sum_{j\in\mathcal{M}_{i}}R_{i,j}\left(t\right)]^{+}+b^{(e,o)}_{i}\left(t\right),
\end{equation}
where $\sum_{j \in \mathcal{M}_{i}} R_{i,j}(t)$ is the total transmission capacity to EFN $i$ in time slot $t$.
The inequality here means that the actual arrival of $Q_{i}^{(e,o)}(t)$ may be less than $b_{i}^{(e,o)}(t)$, because $b^{(e,o)}_{i}(t)$ is the transmission capacity from integrate queue $Q_{i}^{(e,a)}(t)$ to offloading queue $Q_{i}^{(e,o)}(t)$ instead of the amount of truly transmitted workload. 
Recall that we assume the transmission latency from EFT to CFT is negligible compared to the length of each time slot, the workload transmission in each time slot can be accomplished by the end of that time slot.

\subsection{Queueing Model for Central Fog Node}\label{subsec: queue for CFN}

Figure \ref{figure: queue} also shows the queueing model on CFN.
Each CFN $j\in\mathcal{M}$ maintains three queues: 
an arrival queue $Q_{j}^{(c,a)}(t)$, a local processing queue $Q_{j}^{(c,l)}(t)$, and an offloading queue $Q_{j}^{(c,o)}(t)$. Similar to EFNs, workload offloaded from the EFT will be firstly stored in the arrival queue,
then distributed to $Q_{j}^{(c,l)}(t)$ for local processing and to $Q_{j}^{(c,o)}(t)$ for further offloading.


\subsubsection{Arrival Queues in CFNs}

The arrivals on CFN $j$ consist of 
workloads offloaded from EFNs in the set $\mathcal{N}_{j}$. 
We denote the amounts of workloads distributed to the local processing queue and offloading queue in time slot $t$ as $b_{j}^{(c,l)}(t)$ and $b_{j}^{(c,o)}(t)$, respectively, such that
\begin{equation}
	0\leq b^{(c,\beta)}_{j}\left(t\right)\leq b^{(c,\beta)}_{j,\text{max}},\ \forall \beta\in\{l,o\}, \label{constraint: CFN offloading decision 1} 
\end{equation}
where each $b^{(c,\beta)}_{j,\text{max}}$ is a positive constant. Accordingly, $Q_{j}^{(c,a)}(t)$ is updated as follows:
\begin{multline}\label{update: central arrival queue update}
	Q_{j}^{(c,a)}(t+1)\\
	\leq [Q_{j}^{(c,a)}(t)-(b_{j}^{(c,l)}(t)+b_{j}^{(c,o)}(t))]^{+}\!+\!\sum_{i\in\mathcal{N}_{j}}R_{i,j}(t).
\end{multline}

\subsubsection{Offloading Queues in CFNs}

For each CFN $j\in\mathcal{M}$, its offloading queue $Q^{(c,o)}_{j}(t)$ stores the workload to be offloaded to the cloud. 
We define $D_{j}(t)$ as the transmission capacity of the wired link from CFN $j$ to the cloud during time slot $t$, 
which depends on the network state and is upper bounded by some constant $D_{\text{max}}$ for all $j$ and $t$. Then we have the following update function for $Q^{(c,o)}_{j}(t)$:
\begin{equation}\label{update: central offloading queue update}
	Q^{(c,o)}_{j}\left(t+1\right)\leq [Q^{(c,o)}_{j}\left(t\right)-D_{j}\left(t\right)]^{+}+b^{(c,o)}_{j}\left(t\right).
\end{equation}
Note that the amount of actually offloaded workload to the cloud is $\min\{Q_{j}^{(c,o)}(t),D_{j}(t)\}$.

\subsection{Local Processing Queues on EFNs and CFNs}\label{subsec: local queue}

We assume that all fog nodes are able to adjust their CPU frequencies in each time slot, by applying \textit{dynamic voltage and frequency scaling} (DVFS) techniques\cite{mao2017survey}. 
Next, we define $L_{k}^{(\alpha)}$ as the number of CPU cycles that fog node $k\in\mathcal{N}\cup\mathcal{M}$ requires to process one bit of workload, where $\alpha$ is an indicator of fog node $k$'s type ($\alpha=e$ if $k$ is an EFN, and $\alpha=c$ if $k$ is a CFN). 
$L_{k}^{(\alpha)}$ is assumed constant and can be measured offline \cite{miettinen2010energy}. 
Therefore, the local processing capacity of fog node $k$ is $f_{k}^{(\alpha)}(t)/L_{k}^{(\alpha)}$.
The local processing queue on fog node $k$ evolves as follows:
\begin{equation}\label{update: local processing queue update}
	Q_{k}^{(\alpha,l)}(t+1)\!\leq \![Q_{k}^{\left(\alpha,l\right)}\left(t\right)\!-\tau_{0}f_{k}^{\left(\alpha\right)}\left(t\right)\!/\!L_{k}^{\left(\alpha\right)}]^{+}\!+b_{k}^{\left(\alpha,l\right)}\left(t\right).
\end{equation}
All CPU frequencies are nonnegative and finite:
\begin{equation}\label{constraint: CPU frequency}
	0\leq f_{k}^{(\alpha)}\left(t\right)\leq f^{(\alpha)}_{k,\text{max}},\ \forall k\in\mathcal{N}\cup\mathcal{M}\text{ and }t,
\end{equation}
where each $f^{(\alpha)}_{k,\text{max}}$ is a positive constant. 

\subsection{Power Consumptions}\label{subsec: power}

The total power consumptions $P(t)$ of fog tiers in time slot $t$
consist of the processing power consumption and wireless transmit power consumption. 
Given a local CPU with frequency $f$, its power consumption per time slot is  $\tau_{0}\varsigma f^{3}$, where $\varsigma$ is a parameter depending on the deployed hardware and is measurable in practice \cite{kim2018dual}. 
Thus $P(t)$ is defined as follows:
\begin{multline}\label{equation: total fog power consumptions}
	P\left(t\right)\triangleq\hat{P}\left(\boldsymbol{f}\left(t\right),\boldsymbol{p}\left(t\right)\right)=\sum_{i\in\mathcal{N}}\tau_{0}\varsigma(f_{i}^{\left(e\right)}\left(t\right))^{3}\\
	+\sum_{j\in\mathcal{M}}\tau_{0}\varsigma(f_{j}^{\left(c\right)}\left(t\right))^{3}+\sum_{i\in\mathcal{N}}\sum_{j\in\mathcal{M}_{i}} \tau_{0}p_{i,j}\left(t\right),
\end{multline}
where $\boldsymbol{f}(t)\triangleq((f_{i}^{(e)}(t))_{i\in\mathcal{N}},(f_{j}^{(c)}(t))_{j\in\mathcal{M}})$ is the vector of all CPU frequencies, and $\boldsymbol{p}(t)\triangleq(\boldsymbol{p}_{i}(t))_{i\in\mathcal{N}}$ in which $\boldsymbol{p}_{i}(t)=(p_{i,j}(t))_{j\in\mathcal{M}_{i}}$ is the transmit power allocation of EFN $i$. 

\subsection{Problem Formulation}\label{subsec: formulation}

We define the long-term time-average expectation of total power consumptions $\bar{P}
$ and total queue backlog $\bar{Q}$ as follows:
\begin{align}
	&~~~~~~~~~~~~~~\bar{P}\triangleq\limsup_{T\rightarrow\infty}\frac{1}{T}\sum_{t=0}^{T-1}\mathbb{E}\left\{ P\left(t\right)\right\}, \label{definition: time average exp power}\\
	&\bar{Q}\triangleq\limsup_{T\rightarrow\infty}\frac{1}{T}\sum_{t=0}^{T-1}\sum_{\beta\in\{ a,l,o\} }(\sum_{i\in\mathcal{N}}\mathbb{E}\{ Q_{i}^{\left(e,\beta\right)}(t)\}\nonumber \\
	&~~~~~~~~~~~~~~~~~~~~~~~~~~~~~~~~~~~~+\sum_{j\in\mathcal{M}}\mathbb{E}\{ Q_{j}^{(c,\beta)}(t)\}) \label{definition: time average exp backlog}.
\end{align}
In this paper, we aim to minimize the long-term time-average expectation of total power consumptions $\bar{P}$, while ensuring the stability of all queues in the system, \textit{i.e.}, $\bar{Q}<\infty$. 
The problem formulation is given by
\begin{equation}\label{problem: general}
\begin{array}{cl}
\underset{\{\boldsymbol{b}(t),\boldsymbol{f}(t),\boldsymbol{p}(t)\}_{t}}{\text{Minimize}}
	&\displaystyle \bar{P}\\
	\text{Subject to}&\displaystyle (\ref{constraint: EFN offloading decision 1})
	(\ref{constraint: power allocation 1})
	(\ref{constraint: power allocation 2})(\ref{constraint: CFN offloading decision 1})(\ref{constraint: CPU frequency}),\\
	&\displaystyle \bar{Q}<\infty.
\end{array}
\end{equation}


\section{Algorithm Design}\label{sec: algorithm}

\subsection{Predictive Algorithm}

To solve problem (\ref{problem: general}), we adopt Lyapunov optimization techniques\cite{huang2016predictive}\cite{neely2010stochastic} to decouple the problem into a series of subproblems over time slots. We show the detail of this process in Appendix \ref{appendix: pora design}. 
By solving each of these subproblems during each time slot, 
we propose PORA, an efficient and predictive scheme conducts workload offloading in an online and distributed manner. 
We show the pseudocode of PORA in Algorithm \ref{algorithm: pora}. 
Note that symbol $\alpha \in \{ e, c \}$ indicates the type of fog node. 
Specifically, for each fog node $k$, $\alpha=e$ if $k$ is an EFN and CFN otherwise. 
\begin{algorithm}[t]
\caption{Predictive Offloading and Resource Allocation (PORA) in One Time Slot}
\label{algorithm: pora}
\begin{algorithmic}[1]
  \STATE Initialize $\boldsymbol{b}(t)\leftarrow \boldsymbol{0}$, $\boldsymbol{f}(t)\leftarrow\boldsymbol{0}$, $\boldsymbol{p}(t)\leftarrow\boldsymbol{0}$.
  \FOR {each fog node $k\in\mathcal{N}\cup\mathcal{M}$}
    \STATE \%\%\textit{Make Offloading Decisions}
    \IF {$Q_{k}^{(\alpha,a)}(t)> Q_{k}^{(\alpha,l)}(t)$}
      \STATE Set $b_{k}^{(\alpha,l)}(t)\leftarrow b^{(\alpha,l)}_{k,\text{max}}$.
    \ENDIF
    \IF {$Q_{k}^{(\alpha,a)}(t)> Q_{k}^{(\alpha,o)}(t)$}
      \STATE Set $b_{k}^{(\alpha,o)}(t)\leftarrow b^{(\alpha,o)}_{k,\text{max}}$.   
    \ENDIF
    \STATE \%\%\textit{Local CPU Resource Allocation}
    \STATE Set $ f^{(\alpha)}_{k}\left(t\right)\leftarrow\min\{ \sqrt{Q_{k}^{(\alpha,l)}(t)/3V\varsigma L_{k}^{(\alpha)}},f^{(\alpha)}_{k,\text{max}}\}$.
  \ENDFOR
  \STATE \%\%\textit{Transmit Power Allocation}
  \FOR {each EFN $i\in\mathcal{N}$}
    \STATE Set $\lambda_{\text{min}}\leftarrow 0$.
    \STATE Set $\lambda_{\text{max}}\leftarrow \max_{j\in\mathcal{M}_{i}}\frac{(Q_{i}^{(e,o)}-Q_{j}^{(c,a)})H_{i,j}(t)}{N_{0}}-V$.
\WHILE{$\lambda_{\text{max}}-\lambda_{\text{min}}> \varepsilon$}
  \STATE \%\%\textit{Water Filling with Bisection Method}
  \STATE Set $\lambda^{*}\leftarrow (\lambda_{\text{min}}+\lambda_{\text{max}})/2$ 
  \STATE Set $p_{i,j}(t)\!\leftarrow\!B\left[\frac{Q_{i}^{(e,o)}(t)-Q_{j}^{(c,a)}(t)}{V+\lambda^{*}}-\frac{N_{0}}{H_{i,j}(t)}\right]^{+}$.
  \IF {$\sum_{j\in\mathcal{M}_{i}}p_{i,j}(t)> p_{i,\text{max}}$}
      \STATE Set $\lambda_{\text{max}}\leftarrow\lambda^{*}$.
  \ELSE
      \STATE Set $\lambda_{\text{min}}\leftarrow\lambda^{*}$.
  \ENDIF
\ENDWHILE
\ENDFOR
  \STATE Enforce scheduling decisions $\boldsymbol{b}(t)$, $\boldsymbol{f}(t)$, and $\boldsymbol{p}(t)$.
\end{algorithmic}
\end{algorithm}
Next, we introduce PORA in detail.

\subsubsection{Offloading Decision}

In each time slot $t$, under PORA, each fog node $k\in\mathcal{N}\cup\mathcal{M}$ decides the amounts of workload scheduled to the offloading queue and the local processing queue, denoted by $b_{k}^{(\alpha,l)}(t)$ and $b_{k}^{(\alpha,o)}(t)$, respectively. 
Such decisions are obtained by solving the following problem:
\begin{equation}\label{problem: offloading decision}
\begin{array}{cl}
\underset{0\leq b_{k}^{(\alpha,\beta)}\leq b^{(\alpha,\beta)}_{k,\text{max}}}{\text{Minimize}}
	\displaystyle \left(Q^{(\alpha,\beta)}_{k}\left(t\right)-Q_{k}^{(\alpha,a)}\left(t\right)\right)b_{k}^{(\alpha,\beta)},
\end{array}
\end{equation}
where $\beta\in\{l,o\}$. 
Accordingly, the optimal solution to (\ref{problem: offloading decision}) is
\begin{equation}\label{equation: optimal offload decision}
	b_{k}^{(\alpha,\beta)}\left(t\right)=\begin{cases}
b^{(\alpha,\beta)}_{k,\text{max}}, & \text{if }Q_{k}^{(\alpha,\beta)}\left(t\right)<Q_{k}^{(\alpha,a)}(t),\\
0, & \text{otherwise}.
\end{cases}
\end{equation}
From (\ref{equation: optimal offload decision}), 
we see that, to determine the optimal solutions $b_{i}^{(e,l)}(t)$ and $b_{i}^{(e,o)}(t)$, 
each EFN $i$ would compare its integrate queue backlog size $Q_{i}^{(e,a)}(t)$ with its local processing queue backlog size $Q_{i}^{(e,l)}(t)$ and offloading queue backlog size $Q_{i}^{(e,o)}(t)$, respectively.
Particularly, 
if there is too much workload in its integrate queue compared to its local queue ($Q_{i}^{(e,l)}\left(t\right)<Q_{i}^{(e,a)}(t)$),
then it will offload as much workload (up to $b^{(e,l)}_{i,\text{max}}$) as possible to its local queue.  
Likewise, if its integrate queue is loaded with more workload than its offloading queue ($Q_{i}^{(e,o)}\left(t\right)<Q_{i}^{(e,a)}(t)$), it will offload up to an amount of $b^{(e,o)}_{i,\text{max}}$ workload to its offloading queue. 

Notably, if the backlog size of the EFN $i$'s integrate queue is larger than both its local queue and offloading queue, then the EFN will transmit the workload one by one unit (\textit{e.g.} packets); 
each unit of workload is either sent to the EFN $i$'s local queue or its offloading queue, such that the amounts of workload distributed to such two queues are no greater than $b_{i,\text{max}}^{(e,l)}$ and $b_{i,\text{max}}^{(e,o)}$, respectively.
In practice, the workload distributing strategy is left as a degree of freedom to be specified in the implementation of PORA. 
In our simulation, we adopt the following distributing strategy.
When an EFN $i$'s integrate queue backlog size is greater than both its local queue and its offloading backlog size, then it will transmit workload to its local queue until the amount of transmitted workload reaches $b_{i,\text{max}}^{(e,l)}$. 
Then the rest workload in the integrate queue is transmitted to the offloading queue until the amount of distributed workload reaches $b_{i,\text{max}}^{(e,o)}$. 
Such a process terminates whenever the integrate queue becomes empty.

The decision making process is similar for CFNs.
Specifically, each CFN $j$ determines $b_{j}^{(c,l)}(t)$ and $b_{j}^{(c,o)}(t)$ by comparing its arrival queue backlog size $Q_{j}^{(c,a)}(t)$ with its local processing queue backlog size $Q_{j}^{(c,l)}(t)$ and offloading queue backlog size $Q_{j}^{(c,o)}(t)$, respectively.

\textbf{Remark:}
For each EFN, we can view the difference between the backlog sizes of its integrate queue and its local processing/offloading queue as its willingness of workload transmission. 
If such willingness is positive, then the EFN will transmit as much workload as possible from its integrate queue; 
otherwise, the EFN will leave the workload not distributed in the current time slot. 
In such a way, PORA always endeavors to balance the integrate queue backlog and the local/offloading queue backlog.
Likewise, under PORA, each CFN determines its offloading decisions upon the difference between the backlog sizes of its arrival queue and its local processing/offloading queue to ensure the queue stability.
%

\subsubsection{Local CPU Frequency Allocation}
Under PORA, in each time slot $t$, each fog node $k\in\mathcal{N}\cup\mathcal{M}$ sets its local CPU frequency $f_{k}^{(\alpha)}(t)$ by solving the following subproblem:
\begin{equation}\label{problem: CPU frequency}
\begin{array}{cl}
\underset{0\leq f_{k}^{(\alpha)}\leq f^{(\alpha)}_{k,\text{max}}}{\text{Minimize}}
	\displaystyle V\varsigma(f_{k}^{(\alpha)})^{3}-Q_{k}^{\left(\alpha,l\right)}\left(t\right)f_{k}^{(\alpha)}/L_{k}^{(\alpha)}.
\end{array}
\end{equation}
By setting the second derivative of the objective function in (\ref{problem: CPU frequency}) to zero, 
we can obtain the optimal CPU frequency $f_{k}^{(\alpha)}(t)$ to be set by fog node $k$ as
\begin{equation}\label{equation: optimal CPU frequency}
f_{k}^{(\alpha)}(t)=\min\left\{\sqrt{Q_{k}^{\left(\alpha,l\right)}\left(t\right)/3V\varsigma L_{k}^{(\alpha)}},f_{k,\text{max}}^{(\alpha)}\right\}.
\end{equation}
We prove the optimality of (\ref{equation: optimal CPU frequency}) in Appendix \ref{proof: CPU frequency}.

\textbf{Remark:}
When $f^{(\alpha)}_{k}(t) < f_{k,\text{max}}^{(\alpha)}$, the allocated CPU frequency $f_{k}^{(\alpha)}(t)$ is proportional to the square root of the backlog size of local processing queue $Q_{k}^{(\alpha,l)}(t)$ and the inverse of the value of parameter $V$. 
This shows that, on the one hand, PORA would allocate as much CPU frequency as possible to process the workload in the queues. 
On the other hand, the value of parameter $V$ determines the tradeoff between power consumption and the backlog sizes of queues: a small value of $V$ will encourage the fog node to allocate more CPU frequency to process the workload and hence a small queue backlog size; in contrast, a large value of $V$ will make the fog node more conservative to allocate resources, leading to less power consumptions but a large queue backlog size as well. 
In practice, the choice of the value of $V$ is dependent on the system design objective. 

\subsubsection{Power Allocations for EFNs}
In each time slot $t$, under PORA, each EFN $i\in\mathcal{N}$ determines its allocated transmit power $\boldsymbol{p}_{i}(t)$ by solving the following optimization problem.
\begin{equation}\label{problem: power allocation}
\begin{split}
	\underset{\boldsymbol{p}_{i}}{\text{Minimize}}
	&\displaystyle\sum_{j\in\mathcal{M}_{i} }\Big[Vp_{i,j}-m_{i,j}(t)\log_{2}(1+l_{i,j}(t)p_{i,j})\Big]\\
	\text{Subject to}&\displaystyle\sum_{j\in\mathcal{M}_{i}}p_{i,j}\leq p_{i,\text{max}},\\
	 &\displaystyle\ p_{i,j}\geq 0,\ \forall j\in\mathcal{M}_{i},
\end{split}
\end{equation}
where $m_{i,j}(t)\triangleq (Q_{i}^{(e,o)}(t)-Q_{j}^{(c,a)}(t))B$ and $l_{i,j}(t)\triangleq\frac{H_{i,j}(t)}{N_{0}B}$.
By applying \textit{water-filling} algorithm\cite{boyd2004convex}, we obtain the optimal solution to problem (\ref{problem: power allocation}) as
\begin{equation}\label{equation: Pij*}
	p_{i,j}(t)=[m_{i,j}(t)/(V+\lambda^{*})-1/l_{i,j}(t)]^{+},\ \forall j\in\mathcal{M}_{i},
\end{equation}
where $\lambda^{*}$ is the optimal Lagrangian variable that satisfies
\begin{equation}\label{equation: lambda}
	\sum_{j\in\mathcal{M}_{i}}[m_{i,j}\left(t\right)/(V+\lambda^{*})-1/l_{i,j}\left(t\right)]^{+}=p_{i,\text{max}}.
\end{equation} 
The optimality of such solutions is proven in Appendix \ref{proof: transmit power}. 
We adopt bisection method (line 15-25 in Algorithm \ref{algorithm: pora}) to obtain the value of $\lambda^{*}$ with its lower and upper bounds as $\lambda_{\text{min}}$ and $\lambda_{\text{max}}$, respectively. 
Note that the value of $\lambda^{*}$ converges asymptotically to the optimum $\lambda^{\text{opt}}$ as the tolerance parameter $\varepsilon$ approaches zero, such that $|\lambda^{*}-\lambda^{\text{opt}}|\leq\varepsilon/2$.

\textbf{Remark:}
PORA tends to allocate more transmit power to the CFN with smaller arrival queue backlog size $Q_{j}^{(c,a)}(t)$ for load balancing. When $Q_{j}^{(c,a)}(t)\geq Q_{i}^{(e,o)}(t)$, we have $m_{i,j}(t)\leq 0$ and $p_{i,j}(t)=0$, \textit{i.e.}, EFN $i$ allocates no transmit power to CFN $j$ unless the backlog size of the arrival queue on CFN $j$ is greater than that of the offloading queue on EFN $i$.
By increasing the value of $V$, transmit power consumption will be reduced but the backlog size will increase as well.

\subsection{Computational Complexity of PORA}

During each time slot, part of the computational complexity concentrates on the calculation for CPU frequency settings and offloading decision makings. 
Since the calculation (line 3-11) requires only constant time for each fog node, the total complexity of these steps is $O(N+M)$. 
Next, each EFN $i$ applies the bisection method (line 15-26) to calculate the optimal dual variable, with a complexity of $O(\log_{2}((\lambda_{\text{max}}-\lambda_{\text{min}}/\varepsilon)+|\mathcal{M}_{i}|)$. 
After that, EFN $i$ determines the transmit power to each CFN in the set $\mathcal{M}_{i}$. In the worst case, each EFN is potentially connected to all CFNs, thus the total complexity of PORA algorithm is $O(M \times N )$.


\subsection{Performance Analysis}

We conduct theoretical analysis on the relationship between the average power consumption $\bar{P}$ and queue backlog $\bar{Q}$ under PORA scheme in the non-predictive case ($W_{i}=0,\ \forall i\in\mathcal{N}$),
and then analyze the benefits of predictive offloading in terms of latency reduction.

\subsubsection{Time-average Power Consumption and Queue Backlog}

Let $P^{*}$ be the achievable minimum of $\bar{P}$ 
over all feasible non-predictive polices. 
We have the following theorem.
\begin{theorem}\label{theorem: performance}
\textit{Assume the system arrival lies in the interior of the capacity region and $\boldsymbol{Q}(0)<\infty$. Under PORA, without prediction, there exist constants $\theta>0$ and $\epsilon>0$ such that
\begin{equation*}
	\bar{P}\leq \theta/V+P^{*},\ \bar{Q}\leq (\theta+VP_{\text{max}})/\epsilon,
\end{equation*}
where $\bar{P}$ and $\bar{Q}$ are defined in (\ref{definition: time average exp power}) and (\ref{definition: time average exp backlog}), respectively.}
\end{theorem}
The proof is quite standard and hence omitted here.

\textbf{Remark:}
By \textit{Little's} theorem\cite{leon2017probability}, the average queue backlog size is proportional to the average queueing latency. 
Therefore, Theorem \ref{theorem: performance} implies that by adjusting parameter $V$, PORA can achieve an $[O(1/V),O(V)]$ power-latency tradeoff in the non-predictive case. 
Furthermore, the average power consumption $\bar{P}$ approaches the optimum $P^{*}$ asymptotically as the value of $V$ increases to infinity. 

\subsubsection{Latency Reduction}
We analyze the latency reduction induced by PORA under perfect prediction compared to the non-predictive case. 
In particular, we denote the prediction window vector $(W_{i})_{i\in\mathcal{N}}$ by $\boldsymbol{W}$ and the corresponding delay reduction by $\eta(\boldsymbol{W})$. 
For each unit of workload on EFN $i$, let $\pi_{i,w}$ denote the steady-state probability that it experiences a latency of $w$ time slots in $A_{i,-1}(t)$. 
Without prediction, 
the average latency on its \textit{arrival queues} is $
	d=\sum_{i\in\mathcal{N}}\lambda_{i}\sum_{w\geq1}w\pi_{i,w}/\sum_{i\in\mathcal{N}}\lambda_{i}$. Then we have the following theorem.
\begin{theorem}\label{theorem: delay}
\textit{Suppose the system steady-state behavior depends only on the statistical behaviors of the arrivals and service processes. Then the latency reduction $\eta(\boldsymbol{W})$ is
\begin{multline}\label{theorem 2: result 1}
	\eta\left(\boldsymbol{W}\right)\\
	=\frac{\sum_{i\in\mathcal{N}}\lambda_{i}\!\left(\!\sum_{1\leq w\leq W_{i}}\!w\pi_{i,w}\!+\!W_{i}\!\sum_{w\geq1}\!\pi_{i, w+W_{i}}\right)}{\sum_{i\in\mathcal{N}}\lambda_{i}}.
\end{multline}
Furthermore, if $d<\infty$, as $\boldsymbol{W}\rightarrow\infty$, \textit{i.e.}, 
with inifinite predictive information, we have
\begin{equation}\label{theorem 2: result 2}
	\lim_{\boldsymbol{W}\rightarrow\infty}\eta\left(\boldsymbol{W}\right)=d.
\end{equation}
}
\end{theorem}

We relegate the proof of Theorem \ref{theorem: delay} to Appendix \ref{proof: delay}.

\textbf{Remark:}
Theorem \ref{theorem: delay} implies that predictive offloading conduces to a shorter workload latency; 
in other words, with predicted information, PORA can break the barrier of $[O(1/V),O(V)]$ power-latency tradeoff.
Furthermore, the latency reduction induced by PORA is proportional to the inverse of the prediction window size, and approaches zero as prediction window sizes go to infinity. 
In our simulations, we see that PORA can effectively shorten the average arrival queue latency with only mild-value of future information.

\subsection{Impact of Network Topology}

Fog computing systems generally proceed in wireless environments, thus the network topology of such systems is usually dynamic and may change over time slots.
However, at the beginning of each time slot, the network topology is observed and deemed fixed by the end of the time slot. 
Therefore, in the following, we put the focus of our discussion on the impact of network topology within each time slot.

Recall that in our settings, each EFN has access to only a subset of CFNs in its vicinity. For each EFN $i$, the subset of its accessible EFNs is denoted by $\mathcal{M}_{i}$ with a size of $|\mathcal{M}_{i}|$. From the perspective of graph theory, we can view the interconnection among fog nodes of different tiers as a directed graph, in which each vertex corresponds to a fog node and each edge indicates a directed connection between nodes. Hence, the value of $|\mathcal{M}_{i}|$ can be regarded as the out-degree of EFN $i$, which is an important parameter of network topology that measures the number of directed connections originating from EFN $i$. 
Due to time-varying wireless dynamics, the out-degree of each fog node may vary over time slots; 
consequentially, the resulting topology would significantly affect the system performance. In the following, we discuss such impacts under two channel conditions, respectively.

On the one hand, within each time slot, poor channel conditions (\textit{e.g.} in terms of low SINR) would often lead to unreliable or even unavailable connections among fog nodes and hence a network topology with a relatively smaller out-degree of nodes. In this case, each fog node may have a very limited freedom to choose the best target node to offload its workloads, further leading to backlog imbalance among fog nodes or even overloading in its upper tier with a large cumulative queue backlog size. Besides, poor channel conditions may also require more power consumptions to ensure reliable communication between successive fog nodes.

On the other hand, within each time slot, good channel conditions allow each fog node to have a broader access to the fog nodes in its upper tier, resulting a network topology with a relatively larger out-degree of nodes. In this case, each fog node is able to conduct better decision-making with more freedom in choosing the fog nodes in its upper fog tier, thereby achieving a better tradeoff between power consumptions and backlog sizes.

\begin{table}[!t]
\centering
\begin{threeparttable}
\caption{Simulation Settings}
\label{table: simulation parameters}
\begin{tabular}{|c|c|}
\hline
Parameter & Value \\
\hline
$B$ & $2$ MHz \\
\hline
$H_{i,j}(t),\forall i\in\mathcal{N},j\in\mathcal{M}$ & $24\log_{10}d_{i,j}+20\log_{10}5.8$+60 \tnote{a} \\
\hline
$N_{0}$ & $-174$ dBm/Hz \\
\hline
$P_{i,\text{max}},\forall i\in\mathcal{N}$ & $500$ mW \\
\hline
$L^{(e)}_{i}\forall i\in\mathcal{N}$, $L^{(c)}_{j}\forall j\in\mathcal{M}$ & $297.62$ cycles/bit \\
\hline
$f^{(e)}_{i,\text{max}},\forall i\in\mathcal{N}$ & $4$ G cycles/s \\
\hline
$f^{(c)}_{j,\text{max}},\forall j\in\mathcal{M}$ & $8$ G cycles/s \\
\hline
$\varsigma$ & $10^{-27}$ W$\cdot$s$^{3}$/cycle$^{3}$ \\
\hline
$b^{(e,l)}_{i,\text{max}},b^{(e,o)}_{i,\text{max}},\forall i\in\mathcal{N}$ & $6$ Mb/s \\
\hline
$b^{(c,l)}_{j,\text{max}},b^{(c,o)}_{j,\text{max}},\forall j\in\mathcal{M}$ & $12$ Mb/s \\
\hline
$D_{j}(t),\forall j\in\mathcal{M},t$ & $6$ Mb/s \\
\hline
\end{tabular}
\begin{tablenotes}\footnotesize
\item [a] $d_{i,j}$ is the distance between EFN $i$ and CFN $j$.
\end{tablenotes}
\end{threeparttable}
\end{table}

\subsection{Use Cases}
In practice, PORA can be applied as a theoretical framework to design the offloading schemes for fog computing systems under various use cases, such as public safety systems, intelligent transportation, and smart healthcare systems.
For example, in a public safety system, each street is usually deployed with multiple smart cameras (IoT devices). 
At runtime, such smart cameras would upload real-time vision data to one of their accessible EFNs. Each EFN aggregates such data to extract or even analyze the instant road conditions within multiple streets. 
Such EFNs can upload some of the workload to their upper-layered CFNs (each taking charge of one community consisting of several streets) with greater computing capacities. 
Each CFN can further offload the workload to the cloud via optical fiber links. 
For latency-sensitive applications, the real-time vision data will be processed locally on EFNs or offloaded to CFNs. 
For latency-insensitive applications with intensive computation demand, the data will be offloaded to the cloud through the fog nodes. 
PORA conduces to the design of dynamic and online offloading and resource allocation schemes to support such fog systems with various applications.

\section{Numerical Results}\label{sec: simulation}

We conduct extensive simulations to evaluate PORA and its variants.
The parameter settings in our simulation are based on the commonly adopted wireless environment settings that have been used in \cite{liu2017latency, du2017computation}.
The simulation is conducted on a MacBook Pro with 2.3 GHz Intel Core i5 processor and 8GB 2133 MHz LPDDR3 memory, and the simulation program is implemented using Python 3.7.
This section firstly presents the basic settings of our simulations, and then provides the key results under perfect and imperfect prediction, respectively.

\subsection{Basic Settings}

We simulate a hierarchal fog computing system with $80$ EFNs and $20$ CFNs. All EFNs have a uniform prediction window size $W$, which varies from $0$ to $30$. Note that  $W=0$ refers to the case without prediction. 
For each EFN $i$, its accessible CFN set $\mathcal{M}_{i}$ is chosen uniformly randomly from the power set of the CFN set with size $|\mathcal{M}_{i}|=5$. 
We set the time slot length $\tau_{0}=1$ second. 
During each time slot, workload arrives to the system in the unit of packets, each with a fixed size of $4096$ bits. 
The packet arrivals are drawn from previous measurements \cite{benson2010network}, 
where the average flow arrival rate is $538$ flows/s, 
and the distribution of flow size has a mean of $13$ Kb. 
Given these settings, the average arrival rate is about $7$ Mbps. 
All results are averaged over $50000$ time slots. 
We list all other parameter settings in TABLE \ref{table: simulation parameters}.


\begin{figure}[!t]
	\centering
	\includegraphics[width=0.8\linewidth]{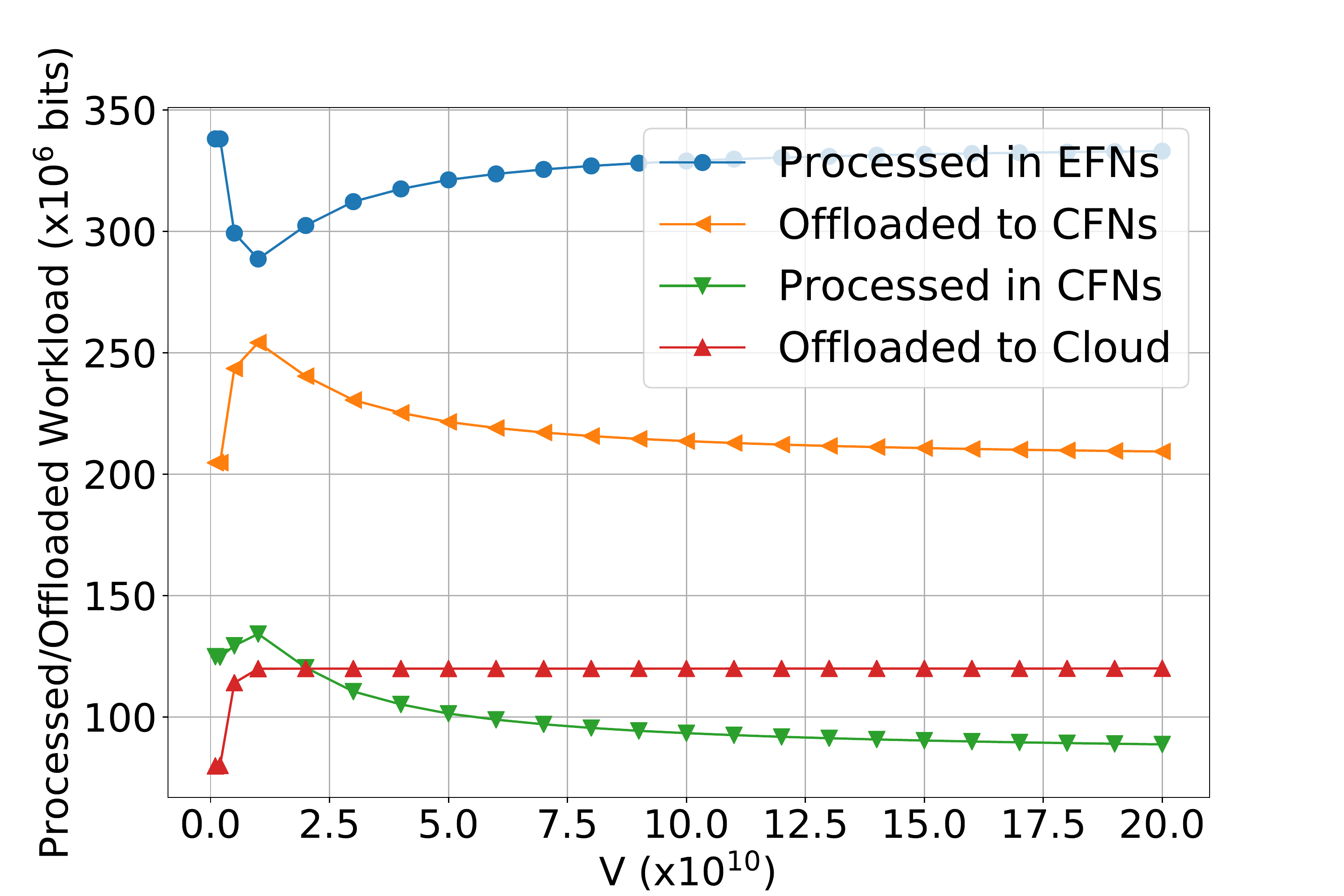}
	\caption{Offloading decisions when $W=10$.}
	\label{figure: offloading decisions vs. V}
\end{figure}

\begin{figure}[!t]
  \centering
  \subfigure[Queue backlogs.]{
    \label{subfig: backlog-W} 
    \includegraphics[width=0.8\linewidth]{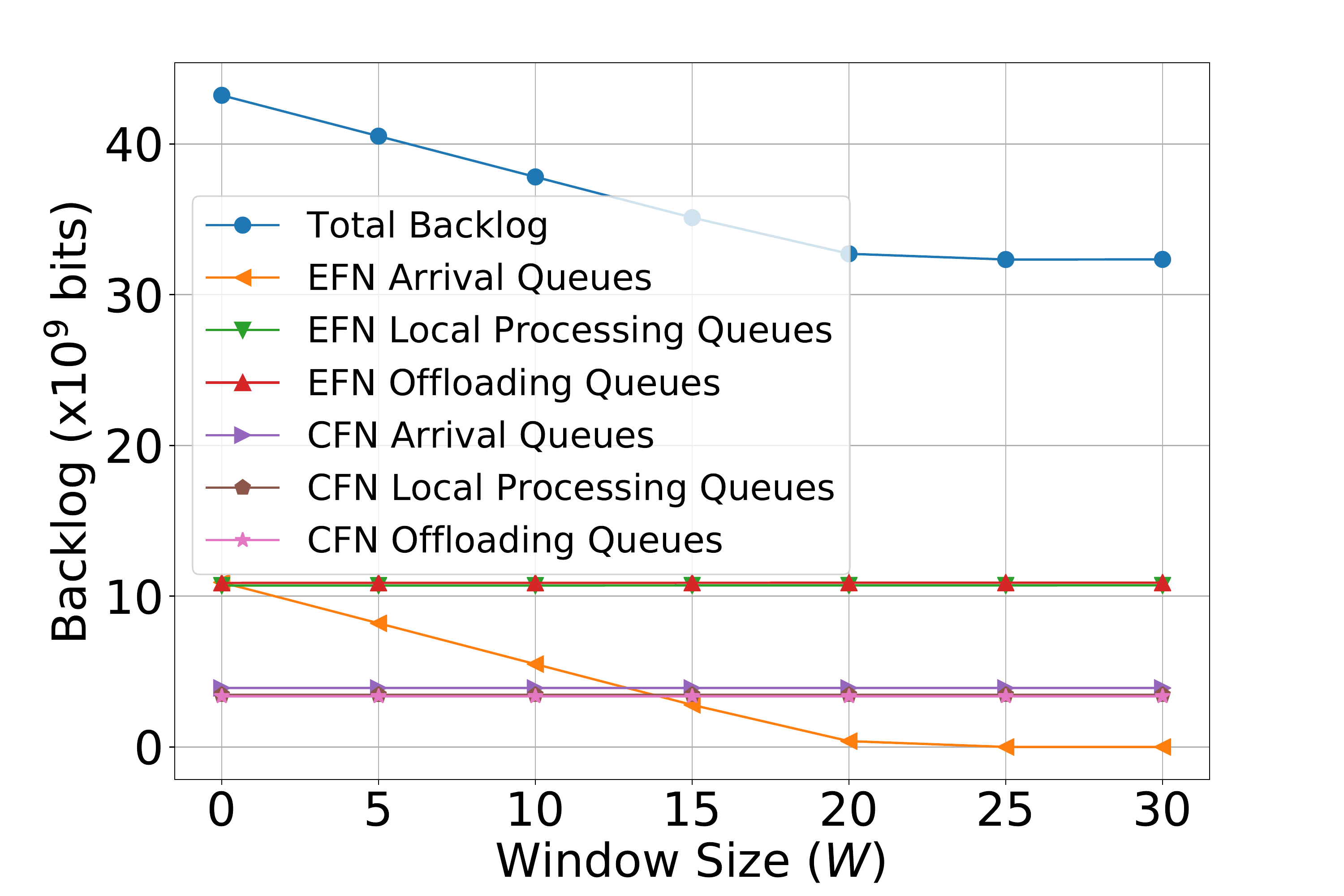}}
  \subfigure[Power consumptions.]{
    \label{subfig: power-W} 
    \includegraphics[width=0.8\linewidth]{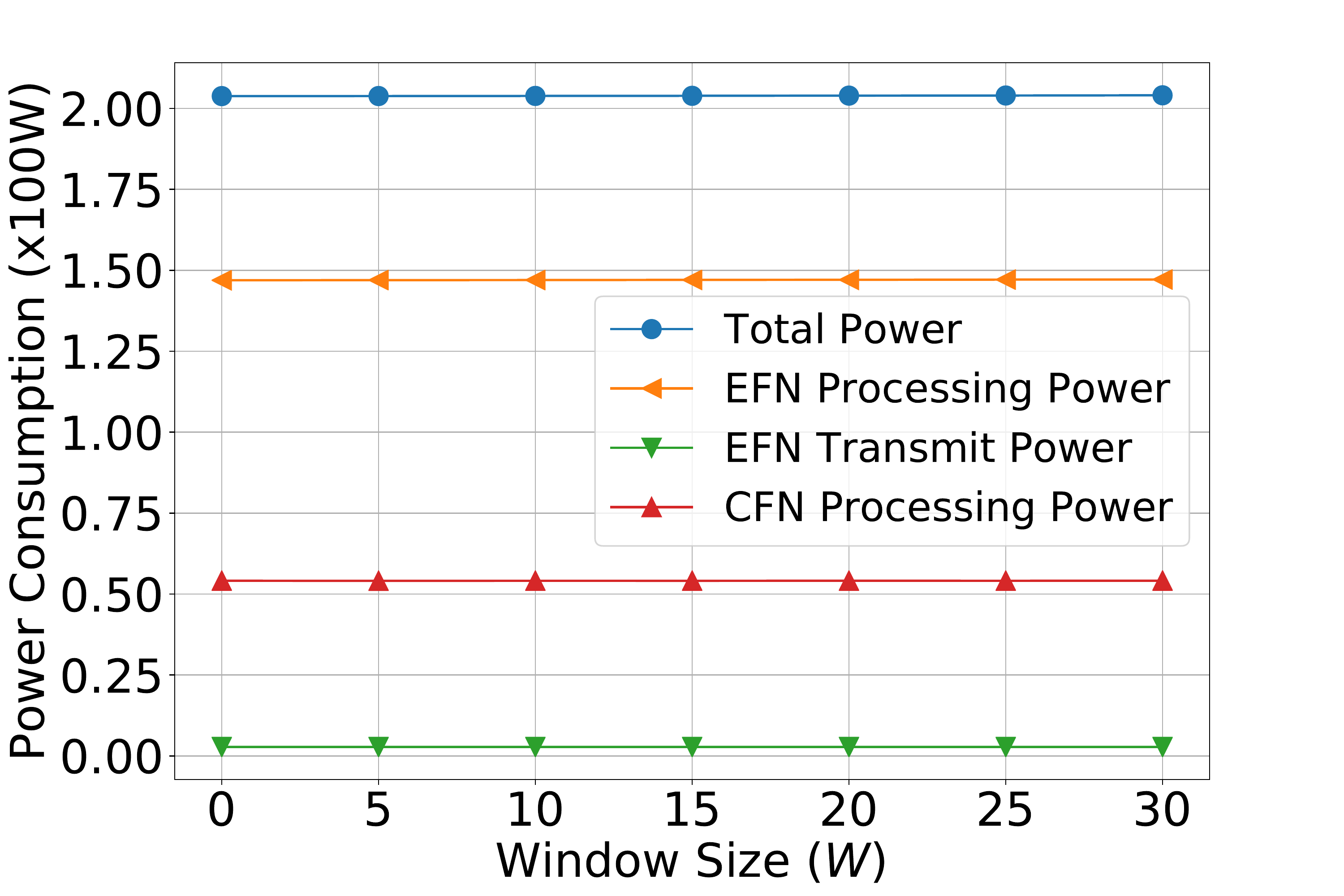}}
  \caption{Performance of PORA vs. $W$ when $V=10^{11}$.}
  \label{figure: performance vs. window size} 
\end{figure}

\subsection{Evaluation with Perfect Prediction}

Under the perfect prediction settings, we evaluate how the values of parameter $V$ and prediction window size $W$ influence the performance of PORA, respectively.

\textbf{System Performance under Different Values of $V$:}
Figure \ref{figure: offloading decisions vs. V} shows the impact of parameter $V$ on the offloading decisions of PORA: 
When the value of $V$ is around $10^{10}$, the time-average amount of locally processed workload on EFNs reaches the bottom of the curve, 
while other offloading decisions induce the peak workload. 
The reason is that the offloading decisions are not only determined by the value of $V$, 
but also influenced by the queue backlog sizes.

\begin{figure}[!t]
  \centering
  \subfigure[Queue backlogs.]{
    \label{subfig: backlog vs. V} 
    \includegraphics[width=0.8\linewidth]{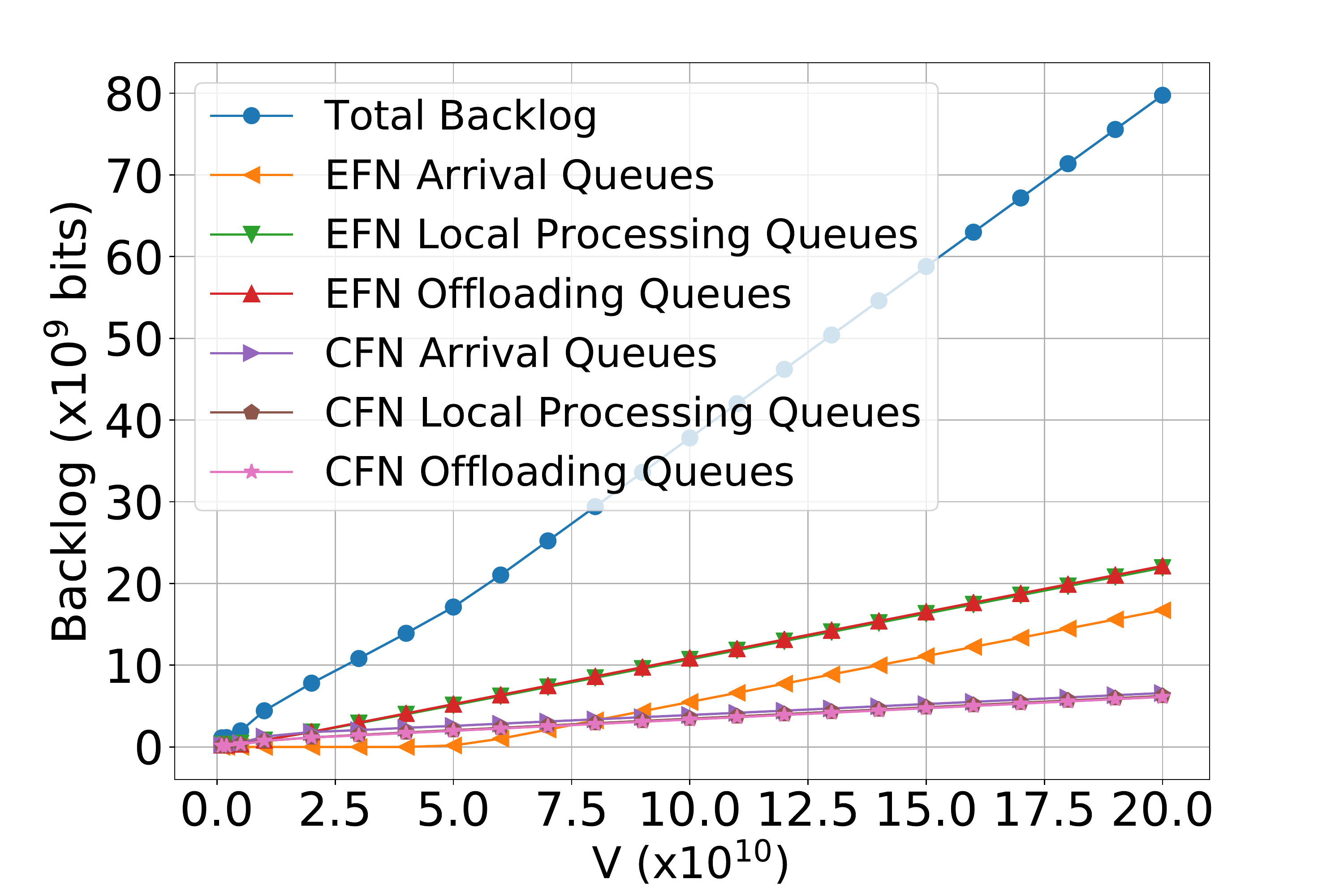}}
  \subfigure[Power consumptions.]{
    \label{subfig: power vs. V} 
    \includegraphics[width=0.8\linewidth]{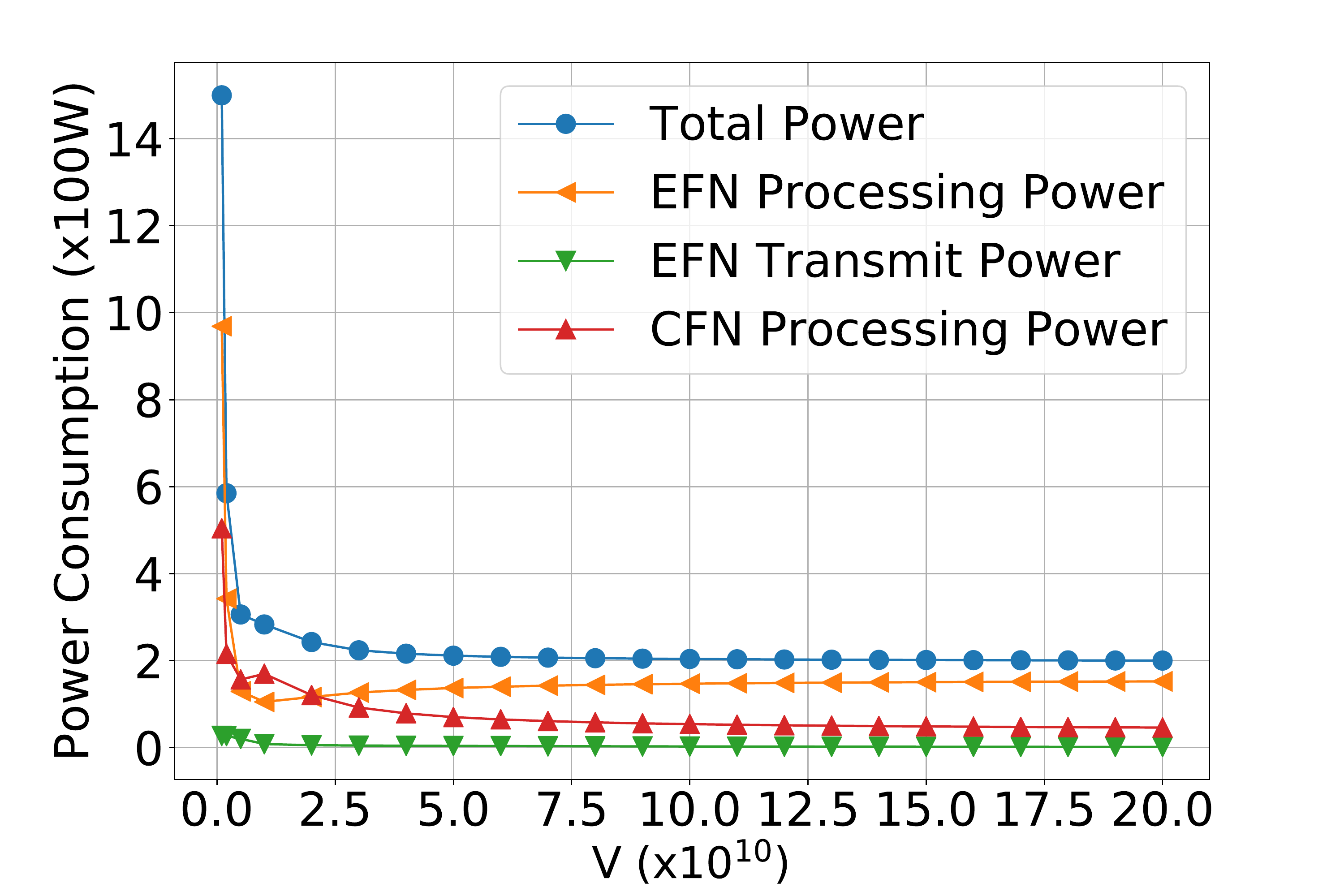}}
  \caption{Performance of PORA when $W=10$.}
  \label{figure: performance vs. V} 
\end{figure}

Figure \ref{figure: performance vs. V} presents the impact of the value of $V$ on different types of queues and power consumptions in the system, respectively. 
As the value of $V$ increases, we see a rising trend in the sizes of all types of queue backlogs, and a roughly falling trend in all types of power consumptions. 


\begin{figure*}[!t]
  \centering
  \subfigure[Total queue backlogs.]{
    \label{subfig: backlog-V-variant} 
    \includegraphics[width=0.4\linewidth]{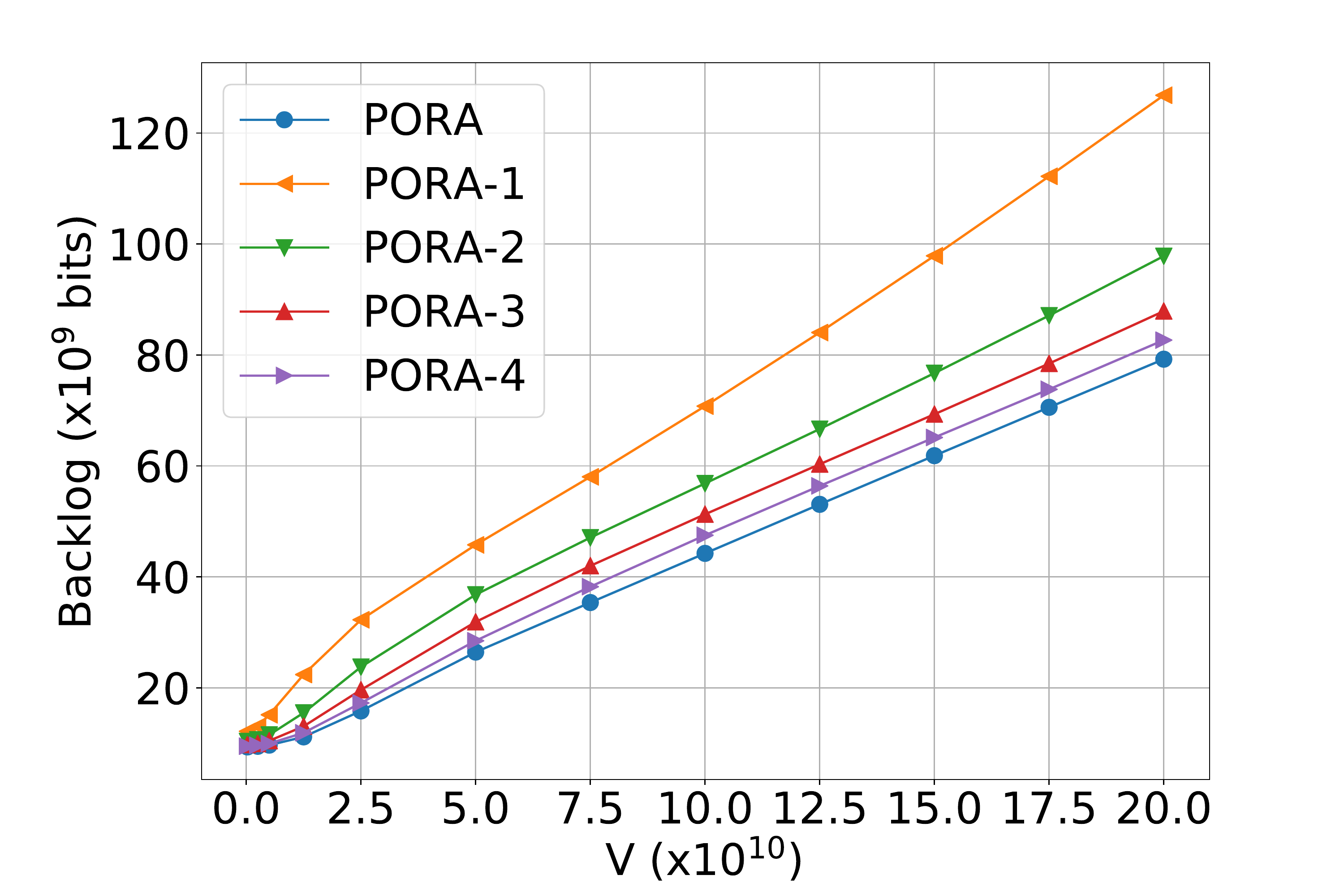}}
  \subfigure[Total power consumptions.]{
    \label{subfig: power-V-variant} 
    \includegraphics[width=0.4\linewidth]{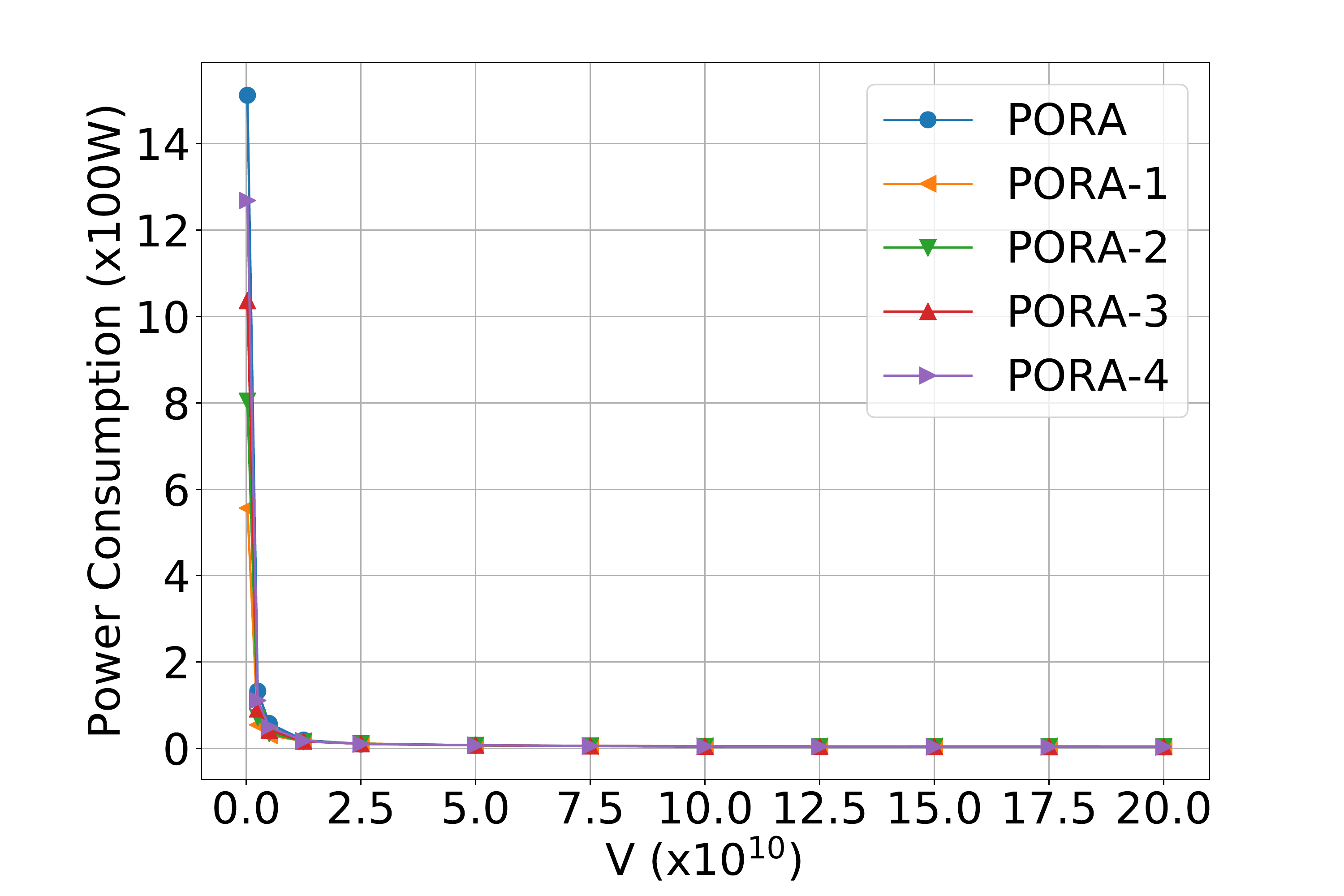}}
  \caption{Performance of variants of PORA.}
  \label{figure: performance v.s. variants} 
\end{figure*}

\begin{figure*}[!t]
  \centering
  \subfigure[Total queue backlogs.]{
    \label{subfig: backlog v.s. time} 
    \includegraphics[width=0.4\linewidth]{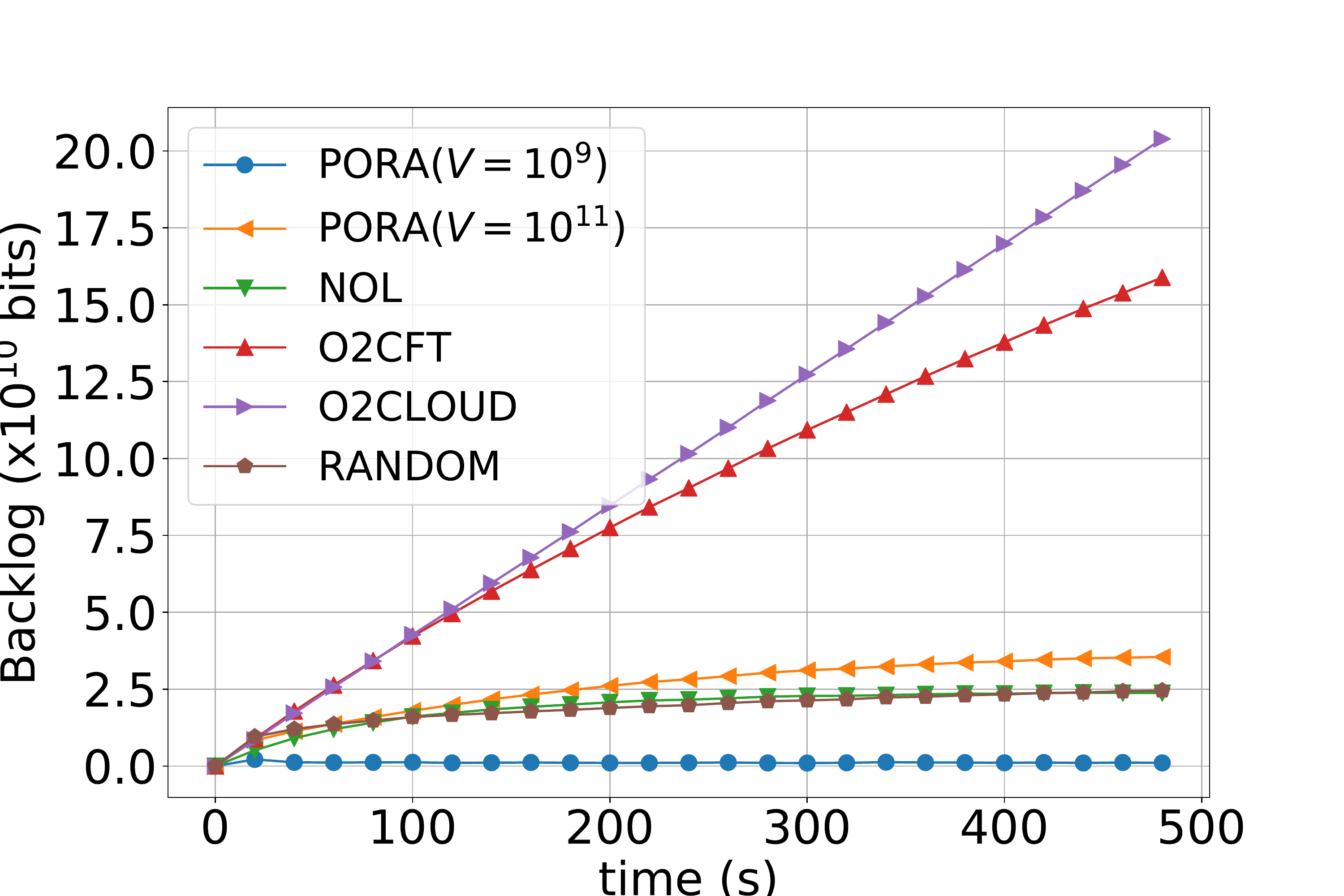}}
  \subfigure[Total power consumptions.]{
    \label{subfig: power v.s. times} 
    \includegraphics[width=0.4\linewidth]{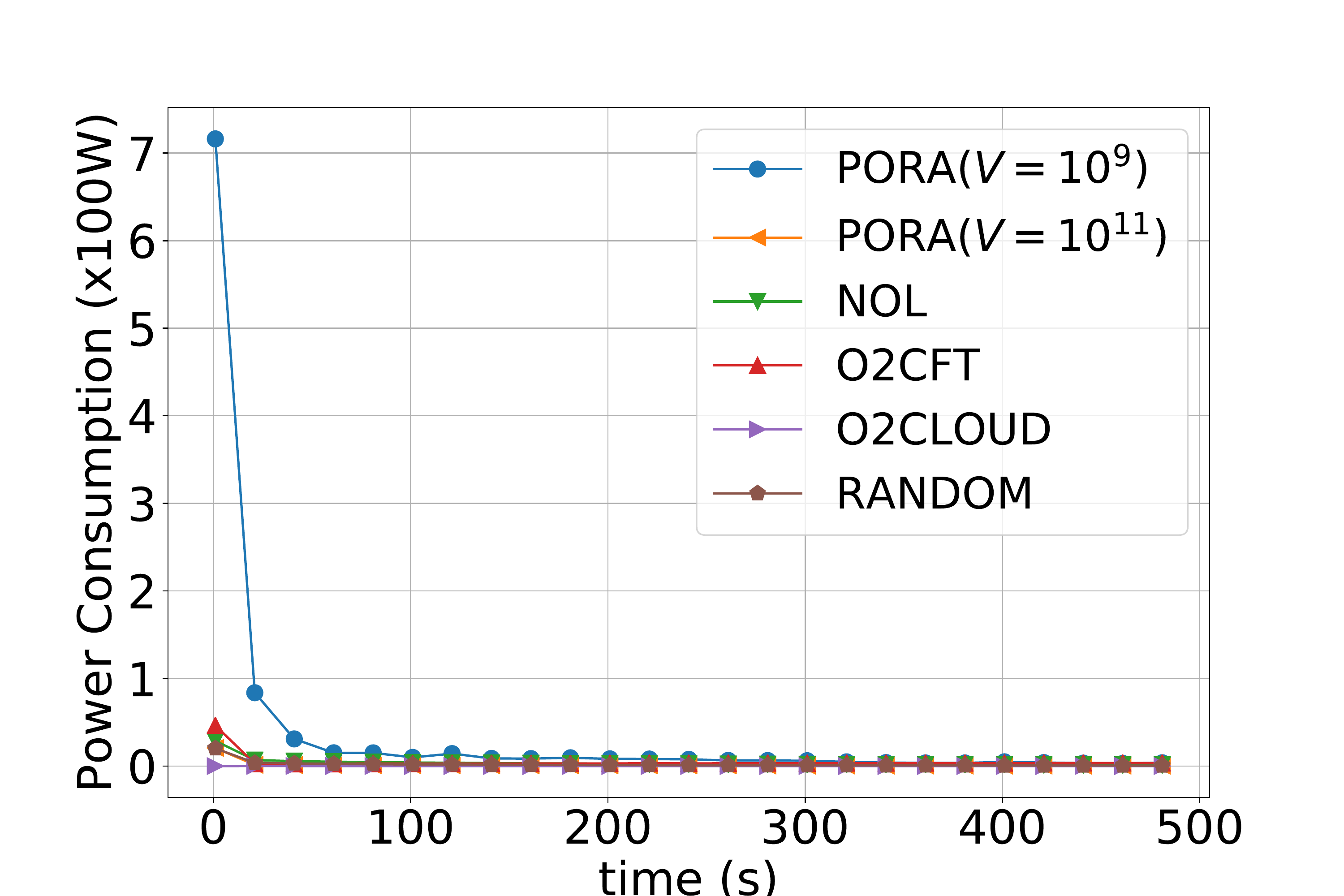}}
  \caption{Comparison between PORA and baselines.}
  \label{figure: comparison} 
\end{figure*}

\textbf{System Performance with Different Values of Prediction Window Size $W$:}
Figures \ref{subfig: backlog-W} and \ref{subfig: power-W} show the system performance with the prediction window size $W$ varying from $0$ to $30$. 
With perfect prediction, PORA effectively shortens the average queueing latencies on EFN arrival queues -- eventually close to zero with no extra power consumption and only a mild-value of prediction window size ($W=20$ in this case).

\textbf{PORA vs. PORA-$d$ (Low-Sampling Variant):}
In practice, since PORA requires to sample system dynamics across various fog nodes, it may incur considerable sampling overheads.
By adopting the idea of randomized load balancing techniques \cite{mitzenmacher2001power}, we propose PORA-$d$, a variant of PORA that reduces the sampling overheads by 
probing $d$ ($d\in\{1,2,3,4\}$) \footnote{When $d=1$, the scheme degenerates to uniform random sampling.} CFNs and conducting resource allocation on which are uniformly chosen for each EFN from its accessible CFN set. 

Figure \ref{figure: performance v.s. variants} compares the performance of PORA with PORA-$d$. 
We observe that PORA achieves the smallest queue backlog size. The result is reasonable since each EFN has access to 5 CFNs under PORA, more than the $d\leq 4$ CFNs under PORA-$d$. As a result, each EFN has more chance to access to the CFNs with better wireless channel condition and processing capacity under PORA when compared with PORA-$d$. The observation that the queue backlog size increases as $d$ decreases further verifies our analysis. In fact, we can view $d$ as the degree of each EFN in the network topology. As $d$ decreases, the system performance degrades.
However, when the value of $V$ is sufficiently large, PORA-$d$ achieves the similar power consumptions as PORA and the ratio of increment in the backlog size is small. For example, when $V=2\times 10^{11}$, PORA-$4$ achieves $4.3$\% larger backlog size than PORA, and PORA-$3$ achieves $10.9$\% larger backlog size than PORA.
In summary, PORA-$d$ (when $d=2,3,4$) can reduce the sampling overheads by trading off only a little performance degradation under large $V$.

\textbf{Comparison of PORA and Baselines:}
We introduce four baselines to evaluate the performance of PORA: (1) NOL (No Offloading): All nodes in the EFT process packets locally. (2) O2CFT (Offload to CFT): All packets are offloaded to the CFT and processed therein. (3) O2CLOUD (Offload to Cloud): All packets are offloaded to the cloud. (4) RANDOM: Each fog node randomly chooses to offload each packet or process it locally with equal chance. 
Note that all above baselines are also assumed capable of pre-serving future workloads in the prediction window.
Figure \ref{figure: comparison} compares the instant total queue backlog sizes and power consumptions over time slots under the five schemes (PORA, NOL, O2CFT, O2CLOUD, RANDOM), 
where $W=10$ and $V\in\{10^{9},10^{11}\}$.

We observe that scheme O2CLOUD achieves the minimum power consumptions, but incurs constantly increasing queue backlog sizes over time. 
The reasons are shown as follows. 
On one hand, in our settings, the mean power consumption for transmitting workload from EFT to CFT is smaller than the mean power consumption of processing the same amount of workload on fog nodes; under scheme O2CLOUD, only wireless transmit power is consumed and hence the minimum is achieved.  
On the other hand, all the workload must travel through all fog tiers before being offloaded to the cloud, which results in network congestion within fog tiers and thus workload accumulation with increasing queue backlogs.

As Figure \ref{figure: comparison} illustrates, PORA achieves the maximum power consumptions but the smallest backlog size when $V=10^{9}$. 
Upon convergence of PORA, the power consumptions under all these schemes reach the same level, but the differences between their queue backlog sizes become more obvious:
PORA ($V=10^{9}$) reduces $96$\% of the queue backlog when compared with NOL and RANDOM.
The results demonstrate that with the appropriate choice of the value of $V$,
PORA can achieve less latency than the four baselines under the same power consumptions.

\subsection{Evaluation with Imperfect Prediction}

In practice, prediction errors are inevitable. 
Hence, we investigate the performance of PORA in the presence of prediction errors \cite{chen2017timely}. 
Particularly, we consider two kinds of prediction errors: false alarm and missed detection.
A packet is falsely alarmed if it is predicted to arrive but it does not arrive actually.
A packet is missed to be detected if it will arrive but is not predicted.
We assume that all EFNs have the uniform false-alarm rate $p_{1}$ and missed-detection rate $p_{2}$.
In our simulation, we consider different pairs of values of $(p_{1},p_{2})$: $(0.0, 0.0)$, $(0.05, 0.05)$, $(0.5, 0.05)$, $(0.05, 0.25)$, and $(0.5, 0.25)$. 
Note that $(p_{1},p_{2})=(0.0,0.0)$ corresponds to the case when the prediction is perfect. 

\begin{figure}[!t]
  \centering
  \subfigure[Total queue backlogs.]{
    \label{subfig: backlog under imperfect prediction} 
    \includegraphics[width=0.8\linewidth]{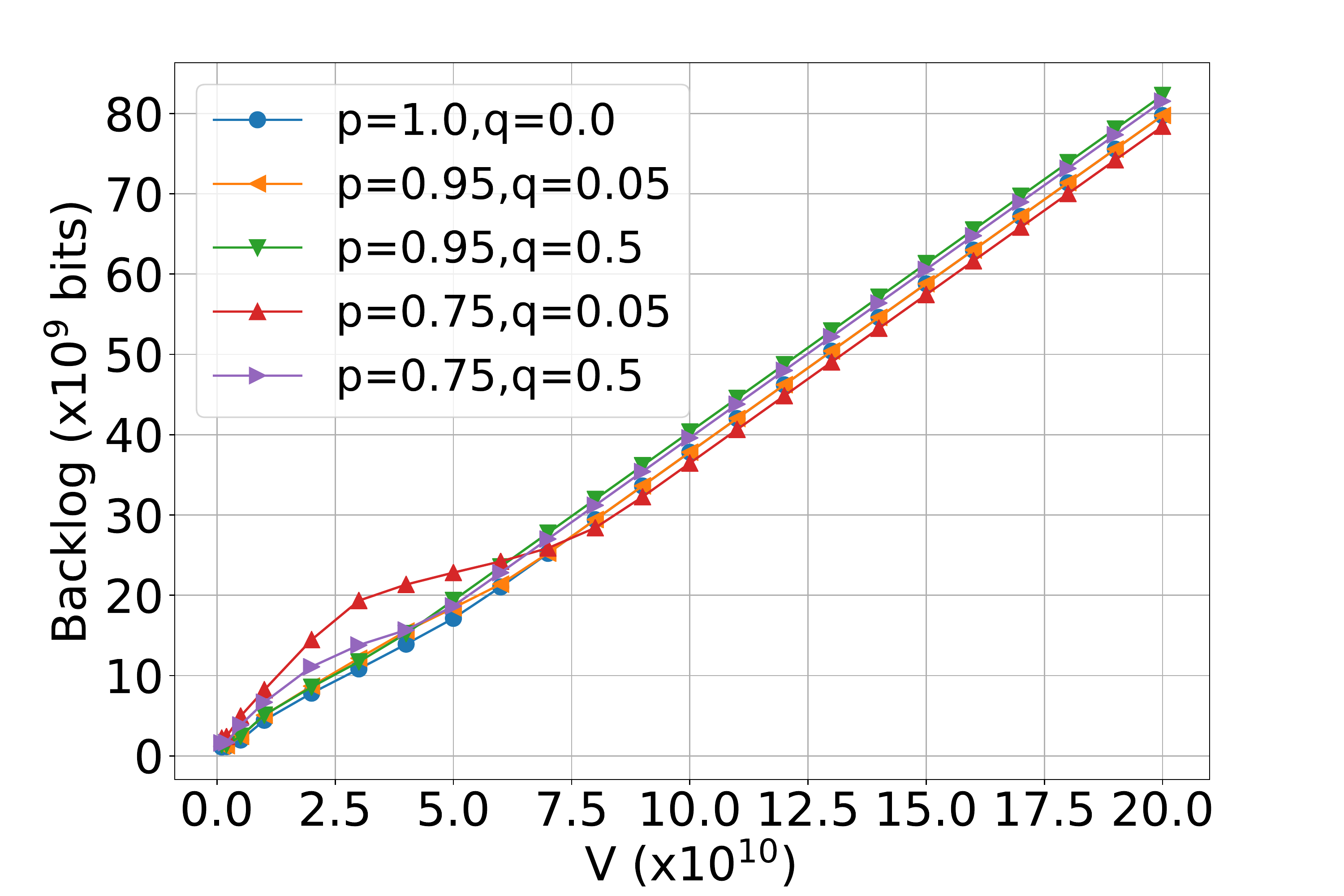}}
  \subfigure[Total power consumptions.]{
    \label{subfig: power under imperfect prediction} 
    \includegraphics[width=0.8\linewidth]{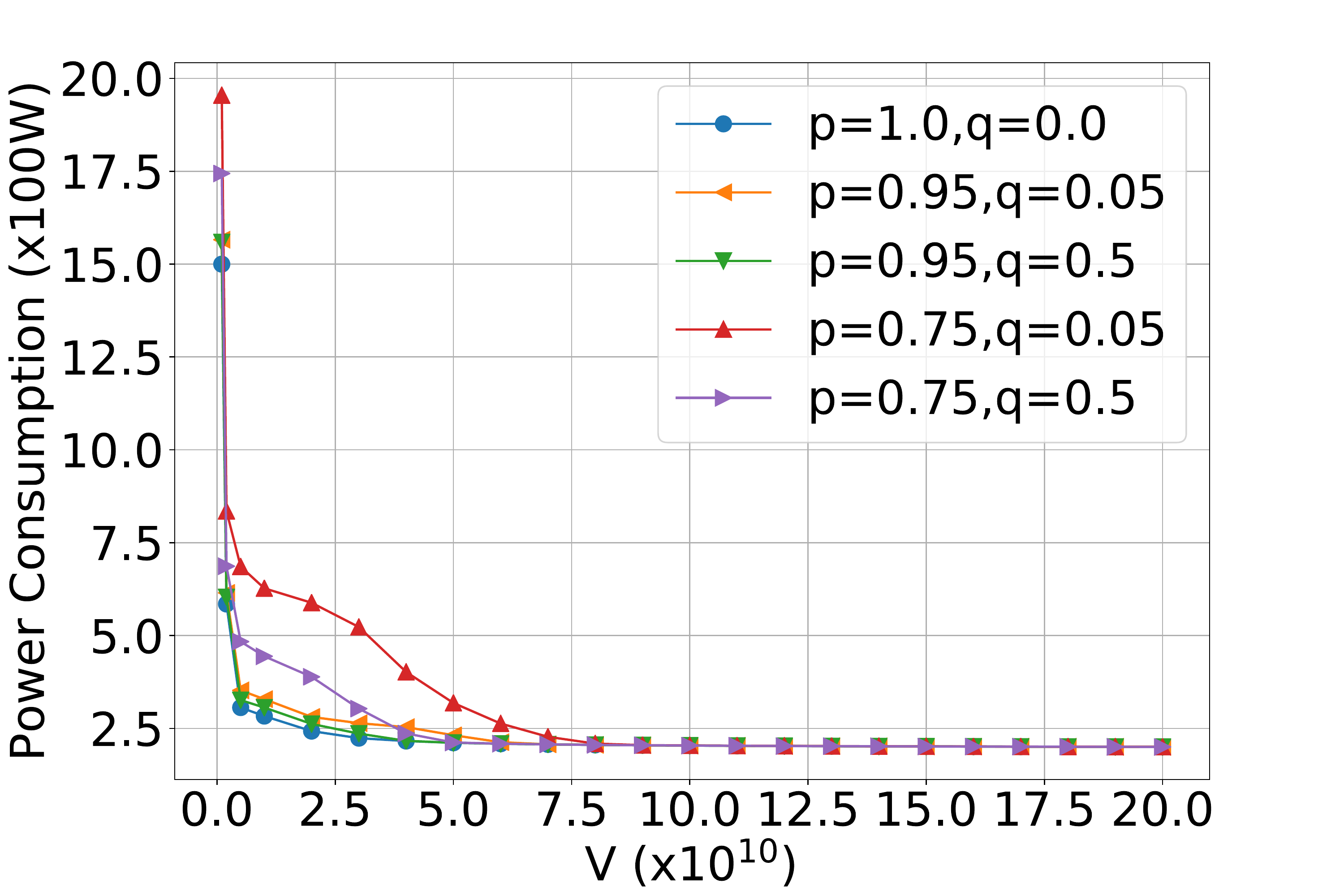}}
  \caption{Performance of PORA under imperfect prediction.}
  \label{figure: performance under imperfect prediction} 
\end{figure}

Figure \ref{figure: performance under imperfect prediction} presents the results under prediction window size $W=10$.
We observe when $V\leq 7.5\times 10^{10}$, both the total queue backlog sizes and power consumptions under imperfect prediction are larger than that under perfect prediction. The reason for this performance degradation is twofold:
First, arrivals that are missed to be detected cannot be pre-served, thus leading to larger queue backlog sizes.
Second, PORA allocates redundant resources to handle the falsely predicted arrivals, thus causing more power consumptions.
As the value of $V$ increases, this performance degradation becomes negligible. Taking the total queue backlog under $(p_{1},p_{2})=(0.25,0.5)$ as an example, when compared with the case under perfect prediction, it increases by $4.72\%$ at $V=10^{11}$, and increases by $2.24\%$ at $V=2\times 10^{11}$. 
Moreover, there is no extra power consumption under imperfect prediction when $V\geq 7.5\times 10^{10}$ since PORA tends to reserve resources to reduce power consumptions under large $V$.

In summary, there will be performance degradation in both total queue backlog sizes and power consumptions in the presence of prediction errors. However, as the value of $V$ increases, this degradation decreases and becomes negligible.
Though a large value of $V$ can improve the robustness of PORA and achieve small power consumptions, it brings long workload latencies. 
In practice, the choice of the value of $V$ depends on how the system designer trades off all these criterions.

\section{Conclusion}\label{sec: conclusion}

In this paper, we studied the problem of dynamic offloading and resource allocation with prediction in a fog computing system with multiple tiers. 
By formulating it as a stochastic network optimization problem, we proposed PORA, an efficient online scheme that exploits predictive offloading to minimize power consumption with queue stability guarantee. 
Our theoretical analysis and trace-driven simulations showed that PORA achieves a tunable power-latency tradeoff, while effectively shortening latency with only mild-value of future information, even in the presence of prediction errors. 
As for future work, our model can be further extended to more general settings such that the instant wireless channel states may be unknown by the moment of decision making or the underlying system dynamics is non-stationary.


%



%
%

\ifCLASSOPTIONcaptionsoff
  \newpage
\fi

\bibliography{references, IEEEabrv}

\appendices

\section{Design of Scheme PORA}\label{appendix: pora design}

First, we define Lyapunov function \cite{neely2010stochastic}  $L\left(\boldsymbol{Q}\left(t\right)\right)$ 
as
\begin{multline}\label{definition: Lyapunov function}
	L(\boldsymbol{Q}(t))\triangleq \frac{1}{2}\sum_{i\in\mathcal{N}}\sum_{\beta\in\{ a,l,o\} }(Q_{i}^{(e,\beta)}(t))^{2} \\
	+\frac{1}{2}\sum_{j\in\mathcal{M}}\sum_{\beta\in\{ a,l,o\} }(Q_{j}^{(c,\beta)}(t))^{2}.
\end{multline}
Next, we define the drift-plus-penalty $\Delta_{V} L\left(\boldsymbol{Q}\left(t\right)\right)$ as
\begin{multline}\label{definition: drift-plus-penalty}
	\Delta_{V}L\left(\boldsymbol{Q}\left(t\right)\right)\triangleq\mathbb{E}\left[L\left(\boldsymbol{Q}\left(t+1\right)\right)-L\left(\boldsymbol{Q}\left(t\right)\right)|\boldsymbol{Q}\left(t\right)\right]\\
	+V\mathbb{E}\left\{ P\left(t\right)|\boldsymbol{Q}\left(t\right)\right\},
\end{multline}
where $V$ is a positive parameter.
According to definition (\ref{definition: Lyapunov function}),
the update functions (\ref{update: prediction sum queue update}), (\ref{update: edge offloading queue update}), (\ref{update: central arrival queue update}), (\ref{update: central offloading queue update}), and (\ref{update: local processing queue update}), there exists a positive constant $\theta>0$ such that
\begin{align}\label{inequality: Lyapunov difference}
	L&\left(\boldsymbol{Q}\left(t+1\right)\right)-L\left(\boldsymbol{Q}\left(t\right)\right)\notag \\
	\leq &\theta\!+\!\sum_{i\in\mathcal{N}}Q_{i}^{(e,a)}(t)\!\left(\!A_{i}(t+W_{i})-b_{i}^{(e,l)}(t)-b_{i}^{(e,o)}(t)\!\right)\!\notag\\
	&+\sum_{i\in\mathcal{N}}Q_{i}^{(e,l)}\left(t\right)\left(b_{i}^{\left(e,l\right)}\left(t\right)-\tau_{0}f_{i}^{\left(e\right)}\left(t\right)/L_{i}^{\left(e\right)}\right)\notag\\
	&+\sum_{i\in\mathcal{N}}Q_{i}^{\left(e,o\right)}\left(t\right)\Big(b_{i}^{\left(e,o\right)}\left(t\right)-\sum_{j\in\mathcal{M}_{i}}R_{i,j}\left(t\right)\Big)\notag\\
	&+\!\sum_{j\in\mathcal{M}}\!Q_{j}^{(c,a)}\left(t\right)\!\bigg[\!\sum_{i\in\mathcal{N}_{j}}\!R_{i,j}\left(t\right)-b_{j}^{(c,l)}\left(t\right)-b_{j}^{(c,o)}\left(t\right)\!\bigg]\notag\\
	&+\sum_{j\in\mathcal{M}}Q_{j}^{(c,l)}\left(t\right)\left(b_{j}^{\left(c,l\right)}\left(t\right)-\tau_{0}f_{j}^{\left(c\right)}\left(t\right)/L_{j}^{\left(c\right)}\right)\notag\\
	&+\sum_{j\in\mathcal{M}}Q_{j}^{(c,o)}\left(t\right)\left(b_{j}^{(c,o)}\left(t\right)-D_{j}\left(t\right)\right).
\end{align}
Substituting (\ref{inequality: Lyapunov difference}) into the definition of drift-plus-penalty shown in (\ref{definition: drift-plus-penalty}) and by $\mathbb{E}[A_{i}(t+W_{i})]=\lambda_{i}$, we obtain
\begin{align}\label{inequality: drift-plus-penalty}
	\Delta&_{V}L\left(\boldsymbol{Q}\left(t\right)\right)\notag \\
	\leq &\theta + V\mathbb{E}\left\{ P\left(t\right)|\boldsymbol{Q}\left(t\right)\right\}\notag  \\
	&+\sum_{i\in\mathcal{N}}Q_{i}^{(e,a)}(t)\mathbb{E}\Big\{\lambda_{i}-\left(b_{i}^{(e,l)}(t)+b_{i}^{(e,o)}(t)\right)|\boldsymbol{Q}(t)\Big\}\notag \\
	&+\sum_{i\in\mathcal{N}}Q_{i}^{(e,l)}\left(t\right)\mathbb{E}\left\{ b_{i}^{\left(e,l\right)}\left(t\right)-\tau_{0}f_{i}^{\left(e\right)}\left(t\right)/L_{i}^{\left(e\right)}|\boldsymbol{Q}\left(t\right)\right\}\notag \\
	&+\sum_{i\in\mathcal{N}}Q_{i}^{\left(e,o\right)}\left(t\right)\mathbb{E}\bigg\{ b_{i}^{\left(e,o\right)}\left(t\right)-\sum_{j\in\mathcal{M}_{i}}R_{i,j}\left(t\right)|\boldsymbol{Q}\left(t\right)\bigg\}\notag \\
	&+\sum_{j\in\mathcal{M}}Q_{j}^{(c,a)}(t)\mathbb{E}\Big\{ \sum_{i\in\mathcal{N}_{j}}R_{i,j}(t)\notag\\
	&-\left(b_{j}^{(c,l)}(t)+b_{j}^{(c,o)}(t)\right)|\boldsymbol{Q}(t)\Big\}\notag \\
	&+\sum_{j\in\mathcal{M}}Q_{j}^{(c,l)}\left(t\right)\mathbb{E}\left\{ b_{j}^{\left(c,l\right)}\left(t\right)-\tau_{0}f_{j}^{\left(c\right)}\left(t\right)/L_{j}^{\left(c\right)}|\boldsymbol{Q}\left(t\right)\right\} \notag\\
	&+\sum_{j\in\mathcal{M}}Q_{j}^{(c,o)}\left(t\right)\mathbb{E}\left\{ b_{j}^{(c,o)}\left(t\right)-D_{j}\left(t\right)|\boldsymbol{Q}\left(t\right)\right\}.
\end{align}
Then by the expression of transmission capacity from EFN $i$ to CFN $j$ shown in (\ref{equation: offload rate to central fog}) and the expression of total power consumptions shown in (\ref{equation: total fog power consumptions}), we have
\begin{align}\label{inequality: drift-plus-penalty upper bound}
	\Delta&_{V}L\left(\boldsymbol{Q}\left(t\right)\right)\notag \\
	\leq &\theta +\sum_{i\in\mathcal{N}}Q_{i}^{(e,a)}\left(t\right)\mathbb{E}\left\{ A_{i}\left(t+W_{i}\right)|\boldsymbol{Q}\left(t\right)\right\}\notag \\
	& +\sum_{i\in\mathcal{N}}\mathbb{E}\left\{ \left(Q_{i}^{\left(e,l\right)}\left(t\right)-Q_{i}^{(e,a)}\left(t\right)\right)b_{i}^{\left(e,l\right)}\left(t\right)|\boldsymbol{Q}\left(t\right)\right\} \notag\\
	& +\sum_{i\in\mathcal{N}}\mathbb{E}\left\{ \left(Q_{i}^{\left(e,o\right)}\left(t\right)-Q_{i}^{(e,a)}\left(t\right)\right)b_{i}^{\left(e,o\right)}\left(t\right)|\boldsymbol{Q}\left(t\right)\right\}\notag \\
	& +\sum_{i\in\mathcal{N}}\mathbb{E}\Bigg\{ V\tau_{0}\varsigma\left(f_{i}^{\left(e\right)}\left(t\right)\right)^{3}-\frac{\tau_{0}Q_{i}^{\left(e,l\right)}\left(t\right)}{L_{i}^{\left(e\right)}}f_{i}^{\left(e\right)}\left(t\right)|\boldsymbol{Q}\left(t\right)\Bigg\}\notag \\
	& +\sum_{j\in\mathcal{M}}\mathbb{E}\left\{ \left(Q_{j}^{\left(c,l\right)}\left(t\right)-Q_{j}^{\left(c,a\right)}\left(t\right)\right)b_{j}^{\left(c,l\right)}\left(t\right)|\boldsymbol{Q}\left(t\right)\right\}\notag \\
	& +\sum_{j\in\mathcal{M}}\mathbb{E}\left\{ \left(Q_{j}^{\left(c,o\right)}\left(t\right)-Q_{j}^{\left(c,a\right)}\left(t\right)\right)b_{j}^{\left(c,o\right)}\left(t\right)|\boldsymbol{Q}\left(t\right)\right\}\notag \\
	& +\sum_{j\in\mathcal{M}}\mathbb{E}\Bigg\{ V\tau_{0}\varsigma\left(f_{j}^{\left(c\right)}\left(t\right)\right)^{3}-\frac{\tau_{0}Q_{j}^{\left(c,l\right)}\left(t\right)}{L_{j}^{\left(c\right)}}f_{j}^{\left(c\right)}\left(t\right)|\boldsymbol{Q}\left(t\right)\Bigg\} \notag\\
	& +\sum_{i\in\mathcal{N}}\sum_{j\in\mathcal{M}_{i}}\mathbb{E}\Big\{ V\tau_{0}p_{i,j}\left(t\right)\notag \\
	&-\tau_{0}m_{i,j}\left(t\right)\log_{2}\left(1+l_{i,j}\left(t\right)p_{i,j}\left(t\right)\right)|\boldsymbol{Q}\left(t\right)\Big\} \notag\\
	& -\sum_{j\in\mathcal{M}}Q_{j}^{\left(c,o\right)}\left(t\right)\mathbb{E}\left\{ D_{j}\left(t\right)|\boldsymbol{Q}\left(t\right)\right\}  
\end{align}
where $m_{i,j}(t)\triangleq(Q_{i}^{(e,o)}(t)-Q_{j}^{(c,a)}(t))B$ and $l_{i,j}(t)\triangleq\frac{H_{i,j}(t)}{N_{0}B}$ for all $i\in\mathcal{N},j\in\mathcal{M}_{i}$.

To solve problem (\ref{problem: general}), we should minimize the upper bound of $\Delta_{V}L\left(\boldsymbol{Q}\left(t\right)\right)$ in every time slot. However, it is hard to solve a minimization problem with expectation. Thus we approximately solve the problem by considering the following deterministic problem in every time slot $t$:
\begin{equation}\label{problem: one time slot}
\begin{split}
	\underset{\boldsymbol{b},\boldsymbol{f},\boldsymbol{p}}{\text{Minimize}}
	&\sum_{i\in\mathcal{N}}\left(Q_{i}^{\left(e,l\right)}\left(t\right)-Q_{i}^{(e,a)}\left(t\right)\right)b_{i}^{\left(e,l\right)}\\
	&+\sum_{i\in\mathcal{N}}\left(Q_{i}^{\left(e,o\right)}\left(t\right)-Q_{i}^{(e,a)}\left(t\right)\right)b_{i}^{\left(e,o\right)}\\
	&+\sum_{i\in\mathcal{N}}\Bigg(V\tau_{0}\varsigma\left(f_{i}^{\left(e\right)}\right)^{3}-\frac{\tau_{0}Q_{i}^{\left(e,l\right)}\left(t\right)}{L_{i}^{\left(e\right)}}f_{i}^{\left(e\right)}\Bigg)\\
	&+\sum_{j\in\mathcal{M}}\left(Q_{j}^{\left(c,l\right)}\left(t\right)-Q_{j}^{\left(c,a\right)}\left(t\right)\right)b_{j}^{\left(c,l\right)}\\
	&+\sum_{j\in\mathcal{M}}\left(Q_{j}^{\left(c,o\right)}\left(t\right)-Q_{j}^{\left(c,a\right)}\left(t\right)\right)b_{j}^{\left(c,o\right)}\\
	&+\sum_{j\in\mathcal{M}}\Bigg(V\tau_{0}\varsigma\left(f_{j}^{\left(c\right)}\right)^{3}-\frac{\tau_{0}Q_{j}^{\left(c,l\right)}\left(t\right)}{L_{j}^{\left(c\right)}}f_{j}^{\left(c\right)}\Bigg)\\
	&+\sum_{i\in\mathcal{N}}\sum_{j\in\mathcal{M}}\Big[V\tau_{0}p_{i,j}-\tau_{0}m_{i,j}\left(t\right)\log_{2}\left(1\right.\\
	&\left.+l_{i,j}\left(t\right)p_{i,j}\right)\Big]\\
	\text{Subject to }&(\ref{constraint: EFN offloading decision 1})(\ref{constraint: power allocation 1})(\ref{constraint: power allocation 2})(\ref{constraint: CFN offloading decision 1})(\ref{constraint: CPU frequency}).
\end{split}
\end{equation} 
Problem (\ref{problem: one time slot}) can be decomposed into  subproblems shown in Section \ref{sec: algorithm}. By solving these subproblems, we develop PORA, an online scheme that indepedently makes predictive offloading decisions $\boldsymbol{b}(t)$, sets CPU frequencies $\boldsymbol{f}(t)$, and allocates transmit powers $\boldsymbol{p}(t)$ in every time slot $t$.
\hfill \IEEEQED

\section{Proof of Optimal Local CPU Frequency}\label{proof: CPU frequency}
To solve the optimal solution to subproblem (\ref{problem: CPU frequency}), we denote its objective function by 
\begin{equation}
	F_{k}^{(\alpha,t)}\left(f_{k}^{(\alpha)}\right)\triangleq V\varsigma\left(f_{k}^{(\alpha)}\right)^{3}-\frac{Q_{k}^{(\alpha,l)}\left(t\right)}{L_{k}^{(\alpha)}}f_{k}^{(\alpha)}.
\end{equation}
Its first- and second-order derivatives are shown as follows:
\begin{align}
	&\frac{dF^{(\alpha,t)}_{k}\left(f_{k}^{(\alpha)}\right)}{d f_{k}^{(\alpha)}}=3V\varsigma \left(f_{k}^{(\alpha)}\right)^{2}-\frac{Q_{k}^{(\alpha,l)}\left(t\right)}{L_{k}^{(\alpha)}},\\
	&~~~~~~~~~~~\frac{d^{2}F^{(\alpha,t)}_{k}\left(f_{k}^{(\alpha)}\right)}{\left(df_{k}^{(\alpha)}\right)^{2}}=6V\varsigma f_{k}^{(\alpha)}.
\end{align}
From the above two derivatives, 
we conclude that function $F^{(\alpha,t)}_{k}(\cdot)$ is convex in interval $[0,f_{k,\text{max}}^{(\alpha)}]$ since its second order derivative satisfies $d^{2}F^{(\alpha,t)}_{k}(\cdot)/(df_{k}^{(\alpha)})^{2}\geq 0$ for $f_{k}^{(\alpha)}\geq 0$. On the other hand, its first order derivative satisfies $dF^{(\alpha,t)}_{k}(\cdot)/df_{k}^{(\alpha)}=0$ when $f_{k}^{(\alpha)}=\sqrt{Q_{k}^{(\alpha,l)}(t)/3V\varsigma L_{k}^{(\alpha)}}$. Thus the minimum point of $F^{(\alpha,t)}_{i}(\cdot)$ over interval $[0,f_{k,\text{max}}^{(\alpha)}]$ is $\min\left\{ \sqrt{Q_{k}^{(\alpha,l)}(t)/3V\varsigma L_{k}^{(\alpha)}},f_{k,\text{max}}^{(\alpha)}\right\} $. 
\hfill \IEEEQED

\section{Proof of Optimal Transmit Power Allocation}\label{proof: transmit power}

We denote the optimal solution to subproblem (\ref{problem: power allocation}) by $\boldsymbol{p}^{*}_{i}(t)$ and the objective function in subproblem (\ref{problem: power allocation}) by $G^{(t)}_{i}\left(\boldsymbol{p}_{i}\right)$. Moreover, we define the following function
\begin{equation}
	G^{(t)}_{i,j}\left(p_{i,j}\right)\triangleq Vp_{i,j}-m_{i,j}\left(t\right)\log_{2}\left(1+l_{i,j}\left(t\right)p_{i,j}\right)
\end{equation}
for each $j\in\mathcal{M}_{i}$. Then $G^{(t)}_{i}\left(\boldsymbol{p}_{i}\right)$ can be expressed as 
\begin{equation}
	G^{(t)}_{i}\left(\boldsymbol{p}_{i}\right)=\sum_{j\in\mathcal{M}_{i}}G^{(t)}_{i,j}\left(p_{i,j}\right).
\end{equation}
We denote the minimizer of function $G^{(t)}_{i,j}(\cdot)$ in interval $[0,\infty)$ by $\tilde{p}_{i,j}^{(t)}$, \textit{i.e.},
\begin{equation}
	\tilde{p}_{i,j}^{\left(t\right)}\triangleq\arg\min_{p_{i,j}\geq0}G_{i,j}^{\left(t\right)}\left(p_{i,j}\right).
\end{equation}

When $m_{i,j}(t)\leq 0$, $G^{(t)}_{i,j}(\cdot)$ is increasing over interval $[0,\infty)$ and $\tilde{p}_{i,j}^{(t)}=0$. In this case, we have $\boldsymbol{p}_{i}^{*}(t)=\tilde{\boldsymbol{p}}^{(t)}_{i}$.
When $m_{i,j}(t)>0$, $G^{(t)}_{i,j}(\cdot)$ is convex in interval $[0,\infty)$ since its second-order derivative satisfies
\begin{equation}
	\frac{d^{2}G_{i,j}^{\left(t\right)}\left(p_{i,j}\right)}{d p_{i,j}^{2}}=\frac{m_{i,j}\left(t\right)\left(l_{i,j}\left(t\right)\right)^{2}}{\left(1+l_{i,j}\left(t\right)p_{i,j}\right)^{2}}>0.
\end{equation}
Thus we obtain $\tilde{p}_{i,j}^{(t)}$ by letting its first-order derivative to be zero: 
\begin{equation}
	\frac{dG_{i,j}^{\left(t\right)}\left(p_{i,j}\right)}{dp_{i,j}}|_{p_{i,j}=\tilde{p}_{i,j}^{(t)}}=V-\frac{m_{i,j}\left(t\right)l_{i,j}\left(t\right)}{1+l_{i,j}\left(t\right)\tilde{p}_{i,j}^{(t)}}=0.
\end{equation}
It follows that when $m_{i,j}(t)>0$,
\begin{equation}\label{equation: reference offloading decision}
\tilde{p}_{i,j}^{\left(t\right)}=\left[\frac{m_{i,j}\left(t\right)}{V}-\frac{1}{l_{i,j}\left(t\right)}\right]^{+}.
\end{equation}
If $\sum_{j\in\mathcal{M}_{i}}\tilde{p}^{(t)}_{i,j}\leq p_{i,\text{max}}$, we have $\boldsymbol{p}_{i}^{*}(t)=\tilde{\boldsymbol{p}}^{(t)}_{i}$ as the constraints in (\ref{problem: power allocation}) are satisfied. Otherwise, we have the following lemma.
\vspace{5mm}
\begin{lemma}\label{lemma: 1}
	If $\sum_{j\in\mathcal{M}_{i}}\tilde{p}^{(t)}_{i,j}> p_{i,\text{max}}$, then $\boldsymbol{p}_{i}^{*}(t)$ must satisfy $\sum_{j\in\mathcal{M}_{i}}p_{i,j}^{*}\left(t\right)=p_{i,\text{max}}$.
\end{lemma}
\vspace{3mm}

\textit{Proof:}
We prove Lemma \ref{lemma: 1} by contradiction. Suppose that there exists $\theta_{1}>0$ such that $\sum_{j\in\mathcal{M}_{i}}p_{i,j}^{*}\left(t\right)+\theta_{1}=p_{i,\text{max}}$. Since $\sum_{j\in\mathcal{M}_{i}}\tilde{p}^{(t)}_{i,j}> p_{i,\text{max}}$, there exist $j'\in\mathcal{M}_{i}$ and $\theta_{2}>0$ such that $p_{i,j'}^{*}(t)<\tilde{p}_{i,j'}^{(t)}-\theta_{2}$. Note that $m_{i,j'}(t)>0$ must hold for $j'$ since $\tilde{p}_{i,j'}^{(t)}>0$.
Now we consider a solution $\boldsymbol{p}_{i}^{0}(t)$ to subproblem (\ref{problem: power allocation}) which satisfies 
\begin{equation}\label{def: p_{i,j}^{0}}
\begin{split}
	&p_{i,j'}^{0}\left(t\right)=p_{i,j'}^{*}(t)+\theta_{3},\\
	&p_{i,j}^{0}\left(t\right)=p_{i,j}^{*}\left(t\right),\ \forall j\in\mathcal{M}_{i}/j',
\end{split}
\end{equation}
where $\theta_{3}\in\left(0,\min(\theta_{1},\theta_{2})\right]$. 
Then $\boldsymbol{p}_{i}^{0}(t)$ is a feasible solution since
\begin{multline}
	\sum_{j\in\mathcal{M}_{i}}p_{i,j}^{0}\left(t\right)=\sum_{j\in\mathcal{M}_{i}}p_{i,j}^{*}\left(t\right)+\theta_{3}\\
	\leq\sum_{j\in\mathcal{M}_{i}}p_{i,j}^{*}\left(t\right)+\theta_{1}=p_{i,\text{max}}.
\end{multline}
By the definition of $p_{i,j'}^{0}(t)$ in (\ref{def: p_{i,j}^{0}}), we have
\begin{equation}
p_{i,j'}^{*}\left(t\right)<p_{i,j'}^{0}\left(t\right)<\tilde{p}_{i,j'}^{\left(t\right)}.	
\end{equation}
Since $G_{i,j'}^{\left(t\right)}\left(\cdot\right)$ is convex and $\tilde{p}_{i,j'}^{\left(t\right)}$ is its unique minimizer, we have
\begin{equation}
	G_{i,j'}^{\left(t\right)}\left(p_{i,j'}^{*}\left(t\right)\right)>G_{i,j'}^{\left(t\right)}\left(p_{i,j'}^{0}\left(t\right)\right)>G_{i,j'}^{\left(t\right)}\left(\tilde{p}_{i,j'}^{\left(t\right)}\right).
\end{equation}
It follows that
\begin{equation}
	G_{i}^{\left(t\right)}\left(\boldsymbol{p}_{i}^{*}\left(t\right)\right)>G_{i}^{\left(t\right)}\left(\boldsymbol{p}_{i}^{0}\left(t\right)\right),
\end{equation}
which contradicts the fact that $\boldsymbol{p}_{i}^{*}\left(t\right)$ is the optimal solution to (\ref{problem: power allocation}). 
Thus $\theta_{1}$ must equal zero and $\boldsymbol{p}_{i}^{*}(t)$ satisfies $\sum_{j\in\mathcal{M}_{i}}p_{i,j}^{*}\left(t\right)=p_{i,\text{max}}$.
\hfill \IEEEQED

\vspace{5mm}

When $\sum_{j\in\mathcal{M}_{i}}\tilde{p}^{(t)}_{i,j}> p_{i,\text{max}}$,
to find the optimal solution to problem (\ref{problem: power allocation}), we need the following lemma as well.
\vspace{5mm}
\begin{lemma}\label{lamma: 2}
	For any $j\in\mathcal{M}_{i}$, if $m_{i,j}\left(t\right)\leq\frac{V}{l_{i,j}\left(t\right)}$, then $p_{i,j}^{*}(t)=\tilde{p}^{(t)}_{i,j}=0$.
\end{lemma}
\vspace{3mm}

\textit{Proof:}
By (\ref{equation: reference offloading decision}), $\tilde{p}_{i,j}^{\left(t\right)}=0$ if and only if $m_{i,j}\left(t\right)\leq\frac{V}{l_{i,j}\left(t\right)}$.
Next, we show that if $m_{i,j}\left(t\right)\leq\frac{V}{l_{i,j}\left(t\right)}$, then the optimal $p^{*}_{i,j}(t)$ must be zero. 
Particularly, we prove it by contradiction.

Assume the optimal $p^{*}_{i,j}(t)>0$, then there must exist a feasible solution $\boldsymbol{p}^{1}_{i}(t)$ 
such that $p_{i,j'}^{1}(t)=p_{i,j'}^{*}\left(t\right)$ for all $j'\in\mathcal{M}_{i}/j$ and $p_{i,j}^{1}(t)=0<p_{i,j}^{*}(t)$. Then we have
\begin{multline}
	G^{(t)}_{i}\left(\boldsymbol{p}_{i}^{*}\left(t\right)\right)-G_{i}^{(t)}\left(\boldsymbol{p}_{i}^{1}\left(t\right)\right)=Vp_{i,j}^{*}\left(t\right)\\
	-m_{i,j}\left(t\right)\log_{2}\left(1+l_{i,j}\left(t\right)p_{i,j}^{*}\left(t\right)\right).
\end{multline}
If $m_{i,j}(t)\leq 0$, according to $p_{i,j}^{*}(t)>0$, we have
\begin{equation}
	G^{(t)}_{i}\left(\boldsymbol{p}_{i}^{*}\left(t\right)\right)-G^{(t)}_{i}\left(\boldsymbol{p}_{i}^{1}\left(t\right)\right)>0.
\end{equation}
If $0<m_{i,j}(t)\leq \frac{V}{l_{i,j}\left(t\right)}$, since $\tilde{p}_{i,j}^{\left(t\right)}\left(t\right)=0<p_{i,j}^{*}\left(t\right)$ is the unique minimizer of $G_{i,j}^{(h)}(\cdot)$ over $[0,\infty)$, we have
\begin{align}
	G^{(t)}_{i}\left(\boldsymbol{p}_{i}^{*}\left(t\right)\right)-G^{(t)}_{i}\left(\boldsymbol{p}_{i}^{1}\left(t\right)\right)
	&=G_{i,j}^{\left(t\right)}\left(p_{i,j}^{*}\left(t\right)\right)\notag\\
	&>G_{i,j}^{\left(t\right)}\left(\tilde{p}_{i,j}^{\left(t\right)}\left(t\right)\right),
\end{align}
which contradicts the fact that $\boldsymbol{p}_{i}^{*}(t)$ is the optimal solution of problem (\ref{problem: power allocation}). Thus for any $j$ with $m_{i,j}\left(t\right)\leq\frac{V}{l_{i,j}\left(t\right)}$, the optimal $p_{i,j}^{*}(t)$ must be zero.	
\hfill \IEEEQED

\vspace{5mm}

We define $\mathcal{M}_{i}^{+}\triangleq\left\{ j|j\in\mathcal{M}_{i},m_{i,j}\left(t\right)>\frac{V}{l_{i,j}\left(t\right)}\right\}$. 
By applying Lemma \ref{lemma: 1} and Lemma \ref{lamma: 2}, when $\sum_{j\in\mathcal{M}_{i}}\tilde{p}^{(t)}_{i,j}> p_{i,\text{max}}$, we just need to solve the following problem:
\begin{equation}\label{problem: power allocation 2}
\begin{split}
	\underset{
	\left(p_{i,j}\right)_{j\in\mathcal{M}_{i}^{+}}
	}{\text{Minimize}}
	&\sum_{j\in\mathcal{M}_{i}^{+}}\Big[Vp_{i,j}-m_{i,j}\left(t\right)\log_{2}\left(1+l_{i,j}\left(t\right)p_{i,j}\right)\Big]\\
	\text{Subject to }&
	\sum_{j\in\mathcal{M}_{i}^{+}}p_{i,j}= P_{i,\text{max}},\\
	&p_{i,j}\geq0,\ \forall j\in\mathcal{M}_{i}^{+}.
\end{split}
\end{equation}
Note that $(p_{i,j}^{*}(t))_{j\in\mathcal{M}_{i}^{+}}$ is the optimal solution to problem (\ref{problem: power allocation 2}) and it satisfies the following KKT conditions:
\begin{align}
	V-\frac{m_{i,j}\left(t\right)l_{i,j}\left(t\right)}{1+l_{i,j}\left(t\right)p_{i,j}^{*}\left(t\right)}&+\lambda^{*}-\mu_{j}^{*}=0,\ \forall j\in\mathcal{M}_{i}^{+}, \label{constraint: KKT1} \\
	\mu_{j}^{*}p_{i,j}^{*}\left(t\right)=&0,\ \forall j\in\mathcal{M}_{i}^{+}, \label{constraint: KKT2} \\
	\lambda^{*},\mu_{j}^{*}\geq0&,\ \forall j\in\mathcal{M}_{i}^{+}, \label{constraint: KKT3} \\
	\sum_{j\in\mathcal{M}_{i}^{+}}p_{i,j}^{*}&\left(t\right)= p_{i,\text{max}}, \label{constraint: KKT4} \\
	p_{i,j}^{*}&\left(t\right)\geq0, \label{constraint: KKT5}
\end{align}
where $\lambda^{*}$ and $(\mu_{j}^{*})_{j\in\mathcal{M}_{i}^{+}}$ are the corresponding optimal dual variables. 
Multiplying both sides of (\ref{constraint: KKT1}) by $p_{i,j}^{*}(t)$, we have
\begin{multline}
	\left(V-\frac{m_{i,j}\left(t\right)l_{i,j}\left(t\right)}{1+l_{i,j}\left(t\right)p_{i,j}^{*}\left(t\right)}+\lambda^{*}\right)p_{i,j}^{*}\left(t\right)\\
	-\mu_{j}^{*}p_{i,j}^{*}\left(t\right)=0.
\end{multline}
It follows by (\ref{constraint: KKT2}) that
\begin{equation}\label{constraint: KKT6}
	\left(V-\frac{m_{i,j}\left(t\right)l_{i,j}\left(t\right)}{1+l_{i,j}\left(t\right)p_{i,j}^{*}\left(t\right)}+\lambda^{*}\right)p_{i,j}^{*}\left(t\right)=0.
\end{equation}
On the other hand, according to (\ref{constraint: KKT1}) and (\ref{constraint: KKT3}), we have
\begin{align}\label{constraint: KKT7}
	\lambda^{*}&=\frac{m_{i,j}\left(t\right)l_{i,j}\left(t\right)}{1+l_{i,j}\left(t\right)p_{i,j}^{*}\left(t\right)}-V+\mu_{i}^{*}\notag\\
	&\geq\frac{m_{i,j}\left(t\right)l_{i,j}\left(t\right)}{1+l_{i,j}\left(t\right)p_{i,j}^{*}\left(t\right)}-V
\end{align}
for every $j\in\mathcal{M}_{i}^{+}$. Now we consider two cases:
\begin{enumerate}
	\item If $\lambda^{*}<m_{i,j}\left(t\right)l_{i,j}\left(t\right)-V$, then (\ref{constraint: KKT7}) holds only if $p_{i,j}^{*}(t)>0$. It follows by  (\ref{constraint: KKT6}) that
	\begin{equation}
		\lambda^{*}=\frac{m_{i,j}\left(t\right)l_{i,j}\left(t\right)}{1+l_{i,j}\left(t\right)p_{i,j}^{*}\left(t\right)}-V,
	\end{equation}
	which yields $p_{i,j}^{*}\left(t\right)=\frac{m_{i,j}\left(t\right)}{V+\lambda^{*}}-\frac{1}{l_{i,j}\left(t\right)}$.
	\item If $\lambda^{*}\geq m_{i,j}(t)l_{i,j}(t)-V$, then condition (\ref{constraint: KKT6}) holds if and only if $p_{i,j}^{*}(t)=0$.
\end{enumerate}
In conclusion, we have
\begin{multline}
	p_{i,j}^{*}\left(t\right)=\\
	\begin{cases}
\frac{m_{i,j}\left(t\right)}{V+\lambda^{*}}-\frac{1}{l_{i,j}\left(t\right)}, & \text{if }\lambda^{*}<m_{i,j}\left(t\right)l_{i,j}\left(t\right)-V, \\
0, & \text{if }\lambda^{*}\geq m_{i,j}\left(t\right)l_{i,j}\left(t\right)-V,
\end{cases}
\end{multline} 
or equivalently,
\begin{equation}\label{equation: Pij* with dual}
	p_{i,j}^{*}\left(t\right)=\left[\frac{m_{i,j}\left(t\right)}{V+\lambda^{*}}-\frac{1}{l_{i,j}\left(t\right)}\right]^{+}.
\end{equation}
Note that the above expression also applies to the case when $m_{i,j}(t)\leq \frac{V}{l_{i,j}(t)}$. 
Then by substituting (\ref{equation: Pij* with dual}) into (\ref{constraint: KKT4}), we obtain 
\begin{equation}
	\sum_{j\in\mathcal{M}_{i}}\left[\frac{m_{i,j}\left(t\right)}{V+\lambda^{*}}-\frac{1}{l_{i,j}\left(t\right)}\right]^{+}=p_{i,\text{max}}.
\end{equation}
The left-hand side is a piecewise-linear decreasing function of $\lambda^{*}$, with the breakpoint at $(m_{i,j}\left(t\right)l_{i,j}\left(t\right)-V)$.
Therefore, the equation has a unique solution.
\hfill \IEEEQED

\section{Proof of Theorem \ref{theorem: delay}}\label{proof: delay}

Applying the Corollary 1 in \cite{huang2016predictive}, given prediction window size $W_{i}$, the average latency of workload in arrival queue $A_{i,-1}(t)$ of EFN $i$ under PORA is 
\begin{equation}
	d_{i}^{p}=\sum_{w\geq1}w\pi_{i,w+W_{i}}.
\end{equation}
According to Little's theorem \cite{leon2017probability}, the average arrival queue backlog size of EFN $i$ under prediction is
\begin{equation}
	\psi_{i}^{p}=\lambda_{i}d_{i}^{p}=\lambda_{i}\sum_{w\geq1}w\pi_{i,w+W_{i}}.
\end{equation}
Therefore, the total average arrival queue backlog sizes of all EFNs is
\begin{equation}\label{equation: average all fog node backlogs with prediciton}
	\psi^{p}=\sum_{i\in\mathcal{N}}\psi_{i}^{p}=\sum_{i\in\mathcal{N}}\lambda_{i}\sum_{w\geq1}w\pi_{i,w+W_{i}}.
\end{equation}
When the prediction window size is zero, \textit{i.e.}, when there is no prediction, the corresponding total average arrival queue backlog size of all EFNs is
\begin{equation}\label{equation: average all fog node backlogs without prediciton}
	\psi=\sum_{i\in\mathcal{N}}\psi_{i}=\sum_{i\in\mathcal{N}}\lambda_{i}\sum_{w\geq1}w\pi_{i,w}.
\end{equation}
Using (\ref{equation: average all fog node backlogs with prediciton}) and (\ref{equation: average all fog node backlogs without prediciton}), we conclude that
\begin{align}
	\psi-\psi^{p}=&\sum_{i\in\mathcal{N}}\lambda_{i}\bigg(\sum_{w\geq1}w\pi_{i,w}-\sum_{w\geq1}w\pi_{i,w+W_{i}}\bigg)\notag \\
	=&\sum_{i\in\mathcal{N}}\lambda_{i}\bigg(\sum_{w\geq1}w\pi_{i,w}-\sum_{w\geq1}\left(w+W_{i}\right)\pi_{i,w+W_{i}}\notag\\
	&+\sum_{w\geq1}W_{i}\pi_{i,w+W_{i}}\bigg)\notag\\
	=&\sum_{i\in\mathcal{N}}\lambda_{i}\bigg(\sum_{w\geq1}w\pi_{i,w}-\sum_{w\geq W_{i}+1}w\pi_{i,w}\notag\\
	&+\sum_{w\geq1}W_{i}\pi_{i,w+W_{i}}\bigg)\notag\\
	=&\sum_{i\in\mathcal{N}}\lambda_{i}\bigg(\!\sum_{1\leq w\leq W_{i}}\!w\pi_{i,w}\!+\!W_{i}\sum_{w\geq1}\!\pi_{i,w+W_{i}}\!\bigg).
\end{align}
Dividing both sides by $\sum_{i\in\mathcal{N}}\lambda_{i}$ and using Little's theorem, we obtain (\ref{theorem 2: result 1}).

Next, we prove (\ref{theorem 2: result 2}). Taking the limit of $\boldsymbol{W}$ ($\boldsymbol{W}\rightarrow\infty$), we obtain
\begin{equation}
	\lim_{\boldsymbol{W}\rightarrow\infty}\sum_{i\in\mathcal{N}}\lambda_{i}\sum_{1\leq w\leq W_{i}}w\pi_{i,w}=\psi.
\end{equation}
It follows that
\begin{equation}\label{equation: limit delay reduction}
	\lim_{\boldsymbol{W}\rightarrow\infty}\!\eta\left(\boldsymbol{W}\right)=d\!+\!\lim_{\boldsymbol{W}\rightarrow\infty}\frac{\sum_{i\in\mathcal{N}}\lambda_{i}W_{i}\sum_{w\geq1}\pi_{i,w+W_{i}}}{\sum_{i\in\mathcal{N}}\lambda_{i}}.
\end{equation}
On the other hand, we have
\begin{align}\label{inequality: limit delay reduction}
	\lim_{\boldsymbol{W}\rightarrow\infty}\eta\left(\boldsymbol{W}\right)&=\frac{\psi}{\sum_{i\in\mathcal{N}}\lambda_{i}}-\lim_{\boldsymbol{W}\rightarrow\infty}\frac{\psi^{p}}{\sum_{i\in\mathcal{N}}\lambda_{i}}\notag\\
	&\leq\frac{\psi}{\sum_{i\in\mathcal{N}}\lambda_{i}}.
\end{align}
Combining (\ref{equation: limit delay reduction}) and (\ref{inequality: limit delay reduction}), we have
\begin{equation}\label{equation: zero}
	\lim_{\boldsymbol{W}\rightarrow\infty}\frac{\sum_{i\in\mathcal{N}}\lambda_{i}W_{i}\sum_{w\geq1}\pi_{i,w+W_{i}}}{\sum_{i\in\mathcal{N}}\lambda_{i}}=0,
\end{equation}
since it cannot be negative. Substituting (\ref{equation: zero}) into (\ref{equation: limit delay reduction}), we obtain
\begin{equation}
	\lim_{\boldsymbol{W}\rightarrow\infty}\eta\left(\boldsymbol{W}\right)=d.
\end{equation}
\hfill \IEEEQED

\end{document}